\documentclass[10pt]{article}
\usepackage[centertags]{amsmath}

\newcommand{\sect}[1]{\setcounter{equation}{0}\section{#1}}

\renewcommand{\theequation}{\thesection.\arabic{equation}}
\def\spazio#1{\vrule height#1em width0em depth#1em}

\usepackage{graphicx}
\usepackage{bm}
\usepackage{graphics}

\def\qa{\Bigl[\Bigr.}
\def\qc{\Bigl.\Bigr]}
\def\ta{\Bigl(\Bigr.}
\def\tc{\Bigl.\Bigr)}

\begin{document}

\title{\bf Two fermion relativistic bound states}

\author{R.Giachetti$\,{}^{1,2}$,
E.Sorace$\,{}^{2}$}

\date{}

\maketitle

\centerline{{\small ${ }^1$ Dipartimento di Fisica, Universit\`a di
Firenze, Italy. }}
\centerline{{\small  ${ }^2$ INFN Sezione di Firenze}}

\begin{abstract}
{
We consider the relativistic quantum mechanics of a two interacting
fermions system. We first present a covariant formulation of
the kinematics of the problem and
give a short outline of the classical results. We then quantize the
system with a general interaction potential and deduce the explicit 
equations in a spherical basis. The case of the Coulomb interaction
is studied in detail by numerical methods, solving the eigenvalue
problem for $j=0$, $j=1$, $j=2$ and determining the spectral curves 
for a varying ratio of the mass of the interacting particles.
Details of the computations, together with a perturbative approach in
the mass ratio and an extended description of the ground states of the
Para- and Orthopositronium are given in Appendix. 
}
\end{abstract}


\bigskip

%
%
%
%


\sect{Introduction}
\label{Sec_introduction}


The relativistic two body problem and the properties of bound states in relativistic
quantum mechanics have a very long history 
and a really large number of solutions have been devised. 
Some of these proposals stress the properties of 
covariance of the model, others assume as
a starting point the constraint dynamics 
or arise directly from a classical symplectic context, 
others determine effective interactions from field theory in 
order to get results to be compared with phenomenological data. The
bound state spectra are considered with particular interest, being 
rather difficult to treat in a pure quantum field theory framework. 
A historical overview  of the subject can be found in \cite{Kr}. 
The research, however, 
has actively continued up to present days and several papers
keep being published both on the mathematical and 
the phenomenological aspects of the problem (see {e.g.} 
\cite{Jun}-\cite{AAVV} and references therein).
 
Some years ago the present authors too gave a contribution
to this discussion, proposing a Lorentz invariant classical 
model for the 
two body Coulomb and gravitational interactions 
where the relativistic covariance is nonlinearly 
realized \cite{GS}.
The paper was written in a plain coordinate language
but it was rather strongly influenced by geometrical and symplectic
ideas. These were developed elsewhere from a mathematical viewpoint and 
in a more general context, using group theoretical ideas for the 
treatment of classical spinning particles, \cite{ST}, or 
studying rigorously the mathematics of  skew-commutative 
phase spaces,  \cite{GRR}. We could however observe that 
a peculiar property of the coordinate system determined in \cite{GS}
-- and briefly recalled in the following Section 2 for the sake
of completeness -- is
that the relative time coordinate disappears from the dynamics,
and a symplectic reduction of
the phase space is made possible, \cite{RG}: 
the relative time plays the role
of a gauge and therefore it can be
given an arbitrary form. The procedure permits then, 
if needed, to reconstruct the two separate world-lines
in the Minkowski space, whose structure and correlation are obviously 
dependent upon the chosen form of the relative time gauge function.

The present paper makes a revisitation of our previous model and 
deals with its canonical quantization, assuming that the 
interacting particles are fermions with arbitrary masses. 
We therefore develop our arguments and we find the
wave equation for the two fermion system in a framework of pure 
relativistic quantum mechanics, without any reference to second 
quantization techniques: this is, probably, the 
simplest way to transfer the classical results to the quantum context
and to obtain a complete control of the symmetry and
covariance properties of the theory. 
The wave equation is determined in the presence of a general scalar
potential, but the explicit results are focused on the Coulomb 
interaction, with the purpose of calculating the spectrum of 
the lowest bound states and in particular of determining the dependence
of the levels upon the mass ratio of the interacting particles
for any value of the total angular momentum.
It proves that the equations and the results are
continuous in the mass ratio, so that they provide a unified
scheme ranging from the Positronium (two equal masses), 
to the Hydrogen atom and finally to the Dirac limit (one of the
masses becomes infinite).  
Although written in a rather compact form,
the equations we find are not easy to study analytically. 
This means that no simple formula for the energy levels is available 
as in the case of the Dirac equation 
and numerical procedures are therefore necessary if we want to determine
the spectroscopy of the system. Appropriate plots will
illustrate the results obtained: from these it will be
possible to establish a regular and simple behavior of the
energy levels {\it vs.} an appropriate mass ratio parameter.
We find, for instance, 
that for two finite masses the degeneracy of the Dirac levels 
$2s_{1/2}\,$-$\,2p_{1/2}$ is removed and for the Hydrogen
atom we reproduce 
a small fraction of the entire Lamb shift:
it is interesting to observe that the correct kinematics
gives a not negligible fraction of the splitting.
The continuous dependence of the spectrum upon the ratio of the
component masses suggests also in a natural way 
a perturbative treatment in the
neighborhood of the Dirac limit. Except for the Positronium, the
main systems of physical interest -- as the Hydrogen atom and, at a
lesser degree, the Muonium and the mesic atom -- have a very small mass
ratio and can be treated accordingly. We have shown how to develop
this framework and we have applied it to the ground state of the
Hydrogen atom.

A comparison with the results of ref. \cite{Chi}-\cite{Dare} is in order. 
Indeed, the equations found in our quantum mechanical framework
agree with those deduced by different methods in the quoted references,
although a bit of work is usually required to reduce the latter.
Concerning the results, we may observe that all those papers
develop a perturbative approach in the fine structure constant up to
$\alpha^4$; complete numerical results are given for equal masses,
\cite{SSM}, and for different masses with a vanishing total angular
momentum, \cite{Dare}. In \cite{Chi} 
we find a perturbative formula, corrected up to $\alpha^4$, expressing 
the Lamb shift due to the Coulomb interaction. The results of
that formula are practically coincident with our numerical results
for the Hydrogen atom, but appear to be less accurate when the masses of
the two fermions become closer and closer: the comparison suggests 
that, at a high degree of accuracy, the perturbation expansions must be
treated with some caution and numerical computations appear
to be necessary and more reliable.
Complete numerical results for equal masses and
values zero and one of the angular momentum, are found in \cite{SSM}.
The computations are carried out by finite element methods and they
are in complete agreement with our results obtained by the 
numerical solution of a singular boundary value problem. 
The authors observe that the Positronium
presents an inversion in the energy of the two levels indicated as 
$2s_{1/2}\,$ and $\,2p_{3/2}$ in the Dirac spectroscopy,
the first being higher than the second. 
As a matter of fact we have drawn the complete plot of these terms
for any mass ratio, showing their actual crossing.
Finally, in the two papers of reference \cite{Dare}, quantum field
theoretical methods are used to determine the equations for fermion
systems under the Coulomb and the complete electromagnetic
interactions respectively. The equations are derived by defining an 
appropriate non conventional vacuum and by studying the conditions 
to be imposed on the coefficient functions of the states 
with a fixed number of particles in order to be eigenstates of the
particle$\,$-$\,$electromagnetic field Hamiltonian. 
The interest is especially addressed to the
properties of the system related to the
variation of the coupling constant $\alpha$
and the perturbative aspects in terms of
$\alpha$ are widely developed. Complete numerical results are given 
for a vanishing value of the angular momentum
and perturbative results up to $\alpha^4$ for non vanishing angular
momentum. The wave equation we derive for the two fermions agrees
with the results of \cite{Dare}, upon restriction to the
static Coulomb interaction. 

The two fermion relativistic bound states are relevant not only
for electrodynamical purposes and atomic physics, but also in the 
framework of phenomenological meson spectroscopy. 
The limited progress of the QCD lattice program for the mesons has
stimulated the interest in new treatments of the two body quantum
systems that could replace the non relativistic Quark Model \cite{IG}
and predict the masses of the mesons, from lighter to the heavier ones,
assuming from the experimental data the least number of values for
physical quark parameters. Numerical approaches have devised new methods
of solving the Bethe-Salpeter equation with interactions derived from
QCD effective potentials, see {\it {e.g.}} \cite{RKM}. New wave
equations from the quantization of constraint theories have been
proposed in a long series of papers, see \cite{CL},
\cite{CVA} and references therein. The authors start from two coupled
Dirac equations; using the Poincar\'e symmetry and some supersymmetric
ideas, they are able to determine the general structure of the
interactions compatible with the constraint scheme. The application to
the phenomenology of mesons gives a very good agreement with the
experimental data.

We have written this paper trying to make it self-consistent. Some
details of the computations have been explained in the Appendixes. 
A brief summary is as follows. 
In Section \ref{Sec_classical} 
we recall the basic definitions and we establish the kinematics
and the dynamics of the system together with their invariance properties
and their validity in any reference frame. In
Section \ref{Sec_quantization} the quantization procedure is 
explained and applied to the
relativistic two fermion problem. In Section \ref{Sec_spectrum} 
we deduce the equations for
an arbitrary scalar potential. Given the general form of the state 
vector with definite energy,
angular momentum and parity -- whose explicit expression involves
eight unknown functions and is reported in Appendix A -- 
we give a short, transparent and comprehensive deduction 
of the equations for the
unknown functions and we specify them to  the particular case
of the Coulomb interaction. The computations leading
to the final result are sketched in Appendix B. 
The Dirac limit is 
treated in Section \ref{Sec_dirac}: 
here we present a procedure that
permits to recover continuously 
the Dirac equation for the finite mass constituent
when the mass of the other goes to infinity. 
Finally, in Section \ref{Sec_results}
we discuss the numerical results with $j=0$, $j=1$ and $j=2$ 
for the whole range of masses: the results for $j=0$
should be compared with  a first order perturbative approach 
in the ratio of the constituent masses, described in Appendix
C. Some few details of the numerical 
procedure are given in Appendix D. In the final Appendix E we
report on some details of the ground state of the Positronium.
In particular we calculate the wave function of the Parapositronium and
Orthopositronium, we show that the non relativistic limit correctly
reproduces the Schr\"odinger equation for the Coulomb problem
and we make some comments on the hyperfine splitting.

In the paper we shall use a unit system with $\hbar=c=1$, and
for the presentation of the results we also assume the electron 
mass equal to unity.


\sect{The classical problem and the dynamical variables}
\label{Sec_classical}


We first outline the procedure for two classical scalar particles
with given masses $m_ {i}$, $i=1\,,\,2$, whose individual Minkowski 
coordinates are denoted by  $x^\mu_{(i)}$ and whose canonical
conjugate momenta are $p^\mu_{(i)}$.
Starting from  the collective coordinates 

\begin{eqnarray}
{}& P^\mu=p^\mu_{(1)}+p^\mu_{(2)}\,,~~~~~~\phantom{X^\mu}
X^\mu=\frac 12 \Bigl(x^\mu_{(1)}+x^\mu_{(2)}\Bigr)\,,\cr
{}\phantom{x}\cr
{}&q^\mu=\frac 12 \Bigl(p^\mu_{(1)}-p^\mu_{(2)}\Bigr)\,,~~~~~~\,
r^\mu\,=x^\mu_{(1)}-x^\mu_{(2)}\,,~~~~~
\end{eqnarray}

\noindent
it has been shown in \cite{GS}, that a good
set of canonical conjugate variables is provided by the following
definitions:
\begin{eqnarray}
&{}& P^\mu,\phantom{XXXXXX}\!Z^\mu=X^\mu+
\displaystyle{\frac
{\varepsilon_{abc}\,P_a\, \eta_b^\mu\, L_c}{\sqrt{P^2}\,[P_0+
\sqrt{P^2}]}}
+\displaystyle{\frac{\varepsilon_a^\mu}{\sqrt{P^2}}}\ta 
q_a\,\breve r-{r_a\,\breve q}\tc+\displaystyle{\frac{P^\mu}{P^2}}\,\breve q\,\breve r
\spazio{1.0}\cr
&{}&\breve{q}=\varepsilon^\mu_0\, q_\mu\,,~~~~~~~\,
\breve{r}=\varepsilon^\mu_0 r_\mu\,,
~~~~~~~~~~~~~~~~~~~~~~~~~~~~~~~~~~
\spazio {1.0}  \cr
&{}&q_a=\varepsilon_a^\mu\, q_\mu\,,~~~~~~
r_a=\varepsilon_a^\mu r_\mu
~~~~~~~~~~~~~~~~~~~~~~~~~~~~~~~~~~
\label{variabili}
\end{eqnarray}

\noindent
where $\eta^{\mu\nu}$ is the Lorentz metric tensor, 
$\varepsilon_{abc}$ is the three dimensional skew-symmetric tensor
and where the sum over repeated indices must be assumed.
\noindent In equation (\ref{variabili}) the tensor
$\,\varepsilon_\alpha^\mu$, $(\mu,\alpha=0,\,3)$  given by
\begin{eqnarray}
&{}&\varepsilon_a^\mu(P)=\eta_a^\mu-\displaystyle{\frac{P_a\,[\,P^\mu
+\eta_0^\mu\sqrt{P^2}\,]}{\sqrt{P^2}\,[\,P_0+\sqrt{P^2}\,]}}\,,
\spazio {1.6} \cr
&{}&\varepsilon_0^\mu(P)=\,P^\mu/\sqrt{P^2}\,
\end{eqnarray}
\noindent realizes the Lorentz transformation to the $P_a=0$
reference frame and therefore it satisfies the identities
\begin{eqnarray}
&{}&\eta_{\mu\nu}\,\varepsilon_\alpha^\mu(P)\,
\varepsilon_\beta^\nu(P)\,=
\,\eta_{\alpha \beta}\,,
\spazio {1.4} \cr
&{}&\eta_{\alpha \beta}\,\varepsilon_\alpha^\mu(P)\,
\varepsilon_\beta^\nu(P)\,=
\,\eta^{\mu\nu}\,.
\end{eqnarray}
\noindent
Indeed it can be observed that both $r_a$ and $q_a$ are Wigner vectors 
of spin
one, as well as
${Z_a}$ has the structure of a position
vector of the Newton-Wigner type for a particle with 
angular momentum ${L_a}$, where
\begin{eqnarray}
L_a=\varepsilon_{abc}\,\,r_b\, q_c\,.
\end{eqnarray}

We can remark that the variables (\ref{variabili}) arise quite
naturally as global and relative coordinates in a two body
Poincar\'e invariant dynamics constructed by using the algebraic
and coalgebraic properties of the Weyl homogeneous spaces. 
Indeed a four component position operator for each
constituent is built in terms of the generators of the corresponding
Weyl algebra, whose non trivial cohomology permits to deduce
global and relative operators. The breaking of the scale invariance 
leaves the resulting dynamical system Poincar\'e symmetric, \cite{ST}.
This procedure, although different in the algebraic assumptions and in
the results, presents clear analogies with the classic paper \cite{BT},
(see also subsequent literature,  {\it{e.g.}} \cite{AAVV}$\,$), where 
the two body theory was for the first time coherently based on the 
algebra of the Poincar\'e generators corresponding to each particle.

Using (\ref{variabili}), the mass shell conditions  
for each of the two particles read
\begin{eqnarray}
{}& p_{(i)}^2=\frac 14\, P^2+(-)^{i+1}\,\sqrt{P^2}\,\breve q +
{\breve q}^2-q_aq_a=m_{i}^2\,. 
\end{eqnarray}
\noindent
from which
\begin{eqnarray}
{}&\sqrt{P^2} \,\,\breve{q}=\frac 12 (m_{1}^2-m_{2}^2)\,,
\cr\spazio {1.4} 
{}&\frac 12 P^2+2\breve{q}^2-2q_a q_a=m_{1}^2+m_{2}^2\,.
\end{eqnarray}
\noindent Therefore the total mass $\lambda=\sqrt{P^2}$ results in
\begin{eqnarray}
{}&  \ta\, q_aq_a+m_{1}^2\,\tc ^{1/2}+\ta 
\,q_aq_a+m_{2}^2\,\tc ^{1/2}\,
=\lambda\,,
\end{eqnarray}
\noindent while the variable $\breve{q}$ can be fixed, generating
a symplectic reduction of the phase space. Its
conjugate coordinate $\breve{r}$ -- the relative time
coordinate -- is cyclic and assumes the character of
a gauge function that is chosen 
 {\it a posteriori} in order to recover the complete Minkowski
description for the two particles. 
In particular cases 
it could be useful to fix $\breve{r}=0$, but in principle
there is no necessity of requiring  such condition. 
 
It is finally straightforward
to construct a Lorentz covariant dynamics by introducing nontrivial 
interaction terms depending only upon $q_a$ and $r_a$. One of the
simplest choices is to add scalar potentials that are functions of the 
Lorentz scalar $~r=(r_ar_a)^{1/2}$. Hence, a relativistic
two-body system interacting by means of a potential $V(r)$ is 
described by 
\begin{eqnarray}
{}& \ta\, q_aq_a+m_{1}^2\,\tc ^{1/2}+\ta \,q_aq_a+m_{2}^2\,
\tc ^{1/2}
=h(r)\,.
\label{P2A}
\end{eqnarray}
with 
\begin{eqnarray}
h(r)=\lambda-V(r)
\label{h(r)}
\end{eqnarray}

The equation of the orbits for the Coulomb interaction, 
$V(r)=-\alpha/r$,
is obtained from the canonical equations 
deduced from (\ref{P2A}). Introducing the variables $u$ and $\theta$
defined by
\begin{eqnarray}
u=\lambda-\displaystyle{\frac\alpha{r}}\,,
~~~~~~~\cos\theta=\displaystyle{\frac{r_3}{r}}\,
\end{eqnarray}
and denoting by $L$ the value of the conserved angular momentum,
the equation reads
\begin{equation}
{\displaystyle\frac{d\theta}{du}}= u\,\qa u^4-(4L^2/\alpha^2)\,u^2\,(\,\lambda-u\,)^2\,
-2(m_1^2+m_2^2)u^2
+(m_1^2-m_2^2)^2\qc^{-1/2} \,(\,4L^2/\alpha^2\,)^{1/2}
\label{orbite}
\end{equation}
and it can be integrated in terms of elliptic functions 
(see \cite{GS} for details). In the case of equal masses,
$m_1=m_2=m$, the
solution for the orbits can be given in terms of elementary functions.
For instance, for $\alpha^2<4L^2$ -- the most relevant case
from our point of view as it gives bounded orbits -- the
solution is
\begin{eqnarray}
{}&\displaystyle{\frac
{4L^2-\alpha^2}{(\,r_ar_a\,)^{1/2}}}=\Bigl( 4L^2\,\lambda^2
+4\,(\alpha^2-4L^2)\,m^2\Bigr)^{1/2}\,
\cos\Bigl(\theta\, [\,1-\alpha^2/4L^2\,]\,^{1/2} \Bigr)-\lambda\,\alpha
\end{eqnarray}
\noindent
and corresponds to the motion of a particle in an external Coulomb
field (see {\it e.g.} \cite{Landau} ).


\sect{The quantization of the free two fermion system}
\label{Sec_quantization}


The next step is the quantization of equation (\ref{P2A})
with $V(r)=0$. The scalar
quantization of the double square root Hamiltonian has received
many attentions and has been discussed also recently in terms of
functional inequalities, \cite{VW}. However, since most of the
physically interesting situations involve fermionic components,
we assume particles of that nature and we quantize the system
accordingly. 

The Dirac operators corresponding to each single particle are
\begin{eqnarray}
&{}& D_1=\Bigl(\frac 12 P_\mu+q_\mu\Bigr)\,
\gamma_{(1)}^\mu-m_1\,,\qquad
D_2=\Bigl(\frac 12 P_\mu-q_\mu\Bigr)\,\gamma_{(2)}^\mu-m_2\,
\label{Dirac1-2}
\end{eqnarray}

\noindent
where we have introduced the following tensor products of gamma matrices
\begin{eqnarray}
\gamma_{(1)}^\mu=\gamma^\mu\otimes{\bf I}_4\,,\qquad
\gamma_{(2)}^\mu={\bf I}_4\otimes\gamma^\mu\,,
\end{eqnarray}
where ${\bf I}_4$ is the unity matrix in four dimensions.

The operators $D_1$ and $D_2$ can be rewritten in terms of the canonical
set (\ref{variabili}). At the same time the constant 
$\gamma_\mu$-matrices 
will be recast in terms of the new set of matrices 
$\{\breve\gamma(P),(\gamma_a(P))_{a=1,3}\}$, where
\begin{eqnarray}
\breve{\gamma}(P)=\varepsilon_0^\mu(P)\gamma_\mu\,,\qquad
\gamma_{a}(P)=\varepsilon_a^\mu(P)\gamma_\mu\,
\label{gammaP}
\end{eqnarray}
The two Dirac equations (\ref{Dirac1-2}) become then 
\begin{eqnarray}
&{}& \frac 12\, \lambda\,\breve{\gamma}_{(1)}+ 
\breve{q}\,\breve{\gamma}_{(1)}
-q_a {\gamma_{(1)}}_{a} = m_1\,,
\qquad
\frac 12\, \lambda\,\breve{\gamma}_{(2)}- 
\breve{q}\,\breve{\gamma}_{(2)}
+q_a {\gamma_{(2)}}_{a} = m_2\,.
\label{Dirac12}
\end{eqnarray}
We observe that the square of $\breve{\gamma}_{(i)}$ gives the unity
matrix. We next 
multiply the equations (\ref{Dirac12}) on the left by 
$\breve{\gamma}_{(1)}$ and $\breve{\gamma}_{(2)}$ 
respectively and we obtain 
\begin{eqnarray}
&{}&\frac 12\, \lambda+ \breve{q}
-q_a \,\breve{\gamma}_{(1)}\,{\gamma_{(1)}}_{a} = 
\,\breve{\gamma}_{(1)}m_1\,,
\qquad
\frac 12\, \lambda- \breve{q}
+q_a \,\breve{\gamma}_{(2)}\,{\gamma_{(2)}}_{a} = 
\,\breve{\gamma}_{(2)}m_2\,.
\end{eqnarray}
By summing and subtracting we finally get 
\begin{subequations}
\label{Diracnew}
\begin{eqnarray}
{}&{}& \lambda=q_a\Bigl( \breve{\gamma}_{(1)}\,{\gamma_{(1)}}_{a}-
\breve{\gamma}_{(2)}\,{\gamma_{(2)}}_{a}\Bigr)+\breve{\gamma}_{(1)}
m_1+
\breve{\gamma}_{(2)}m_2
\label{Diracnew1}\spazio{1.4}\\
{}&{}& \breve{q}= {\frac {1}{2}}\, q_a\Bigl( \breve{\gamma}_{(1)}\,{\gamma_{(1)}}_{a}+
\breve{\gamma}_{(2)}\,{\gamma_{(2)}}_{a}\Bigr)+\breve{\gamma}_{(1)}
m_1-
\breve{\gamma}_{(2)}m_2
\label{Diracnew2}
\end{eqnarray}
\end{subequations}
The right hand sides of the two equations 
(\ref{Diracnew}) are commuting. Multiplying those equations 
term by term, we get 
\begin{eqnarray}
\lambda\,\breve{q}=m_1^2-m_2^2\,.
\end{eqnarray}
and we see that the variable $\breve{q}$ remains fixed, in
complete agreement with the classical symplectic reduction. In fact the
canonical conjugate relative time coordinate  $\breve{r}$ is 
cyclic again: this is signified, in particular, by 
the Lorentz scalar identity for the phase of plane waves 
\begin{eqnarray}
{p_{(1)}}^\mu\, {x_{(1)}}_\mu+{p_{(2)}}^\mu \,{x_{(2)}}_\mu
=P^\mu Z_\mu-q_a\,r_a\,.
\end{eqnarray}
We finally observe that the definition (\ref{gammaP}) is actually a
Lorentz transformation on the $\,\gamma^\mu\,$ four vector
determining a unitary transformation. Therefore, as long as 
$P$ is conserved, the matrices (\ref{gammaP}) will be represented by the 
usual $\gamma$ matrices.
The different behavior under Lorentz transformations is however
signified by the use of the notation (\ref{gammaP}). 
In conclusion equation (\ref{Diracnew1}) is  the
Lorentz-invariant equation for the two-fermion system. 
It is also straightforward to calculate the explicit 
expressions for the sixteen eigenvalues $\lambda$ and the result 
is rather obvious. In fact we have four singular values
\begin{eqnarray}
&{}&\lambda=\,\pm\ta{q_a\,q_a+m_1^2}\tc^{1/2}
\pm\ta{q_a\,q_a+m_2^2}\tc^{1/2}
\,,
\spazio {1.4} \cr
&{}&\phantom{\lambda=}\, 
\pm\ta{q_a\,q_a+m_1^2}\tc^{1/2}
\mp\ta{q_a\,q_a+m_2^2}\tc^{1/2}
\,,
\spazio{1.6}
\label{singval}
\end{eqnarray}
each one with multiplicity four.
Defining
\begin{eqnarray}
M=m_1+m_2\,,\qquad \mu=m_1-m_2\,, \qquad  \rho=\mu/M\,,
\label{masse}
\end{eqnarray}
we make a linear transformation on the tensor product of the two spinor
spaces  that diagonalizes the system at rest,
in such a way that the four singular values (\ref{singval})  
appear in the
order $M$, $-M$, $-\mu$ and $\mu$.
We find it also convenient to perform a further linear transformation 
that, in each four dimensional eigenspace, diagonalizes the square
and  the third  component of the total spin
\begin{eqnarray}
{\bm S}={\bf I}_4\otimes {\bm \sigma} + 
{\bm \sigma}\otimes {\bf I}_4
\label{spintot}
\end{eqnarray}
${\bm \sigma}$ being the Dirac spin. For each singular value of the
mass, the diagonalized spin
will then be ordered with the triplet always following the singlet.

The wave equation will now be obtained by applying the canonical
quantization rules to the equation (\ref{Diracnew1}). 
In view of the bases we have chosen, the natural
coordinates to be used are the spherical ones, in terms of which   
the free Hamiltonian operator reads
\begin{eqnarray}
H_0 =\left(\,
\begin{matrix}
 {\cal J}_M & {\cal H}_0 \spazio{0.8}\cr
 {\cal H}_0  & {\cal J}_\mu\cr
\end{matrix}
\,\right) \,,
\label{H}
\end{eqnarray}
where each matrix element is actually an $8\times 8$ block. Explicitly
\begin{eqnarray}
{\cal J}_M=M\,\left(\,
\begin{matrix}
{\bf I}_4 & 0 \spazio{0.8}\cr
0 & - {\bf I}_4 
\end{matrix}
\,\right)\,,\qquad\qquad
{\cal J}_\mu=\mu\,\left(\,
\begin{matrix}
{\bf I}_4 & 0\spazio{0.8} \cr
0 & - {\bf I}_4 
\end{matrix}
\,\right)
\end{eqnarray}
and
\begin{eqnarray}
{\cal H}_0 =\left(\,
\begin{matrix}
\phantom{-}0&  \phantom{-}X_+&  \phantom{-}X_0&  
\phantom{-}X_-&  \phantom{-}0&  \phantom{-}X_+& 
 \phantom{-}X_0&  \phantom{-}X_- 
 \spazio{0.8}\cr
 - X_-&  \phantom{-}X_0&  \phantom{-}X_-&  \phantom{-}0&  
  - X_-&  
 - X_0&   - X_- &  \phantom{-}0 
 \spazio{0.8}\cr
\phantom{-}X_0&   - X_+&  \phantom{-}0&  \phantom{-}X_-&  
\phantom{-}X_0&  \phantom{-}{X
_{+}}&  \phantom{-}0&   - X_- 
\spazio{0.8}\cr
 - X_+&  \phantom{-}0&   - X_+&   - X_0&   - X_+
&  \phantom{-}0&  \phantom{-}X_+&  \phantom{-}X_0 
\spazio{0.8}\cr
\phantom{-}0&  \phantom{-}X_+&  \phantom{-}X_0&  
\phantom{-}X_-&  \phantom{-}0&  \phantom{-}X_+& 
 \phantom{-}X_0&  \phantom{-}X_-  
 \spazio{0.8}\cr
 - X_-&   - X_0&   - X_-&  0&   - X_-
&  \phantom{-}X_0&  \phantom{-}X_- &  \phantom{-}0 
\spazio{0.8}\cr
\phantom{-}X_0&  \phantom{-}X_+&  \phantom{-}0&  
 - X_-&  \phantom{-}X_0&   - 
X_+&  \phantom{-}0&  \phantom{-}X_- 
\spazio{0.8}\cr
 - X_+&  \phantom{-}0&  \phantom{-}X_+&  \phantom{-}X_0&   
 - X_+&    
\phantom{-}0&   - X_+&   - X_0
\end{matrix}
\,\right)
\label{Xi}
\end{eqnarray}
\medskip

\noindent
The matrix elements in (\ref{Xi}) are the spherical operators
\begin{eqnarray}
X_\pm=-2^{-1/2} \,( \, \pm{q_{x}} + i\,{q
_{y}}\,)\,, \qquad
X_0={q_{z}}\,,
\end{eqnarray}
where $\,q_a \rightarrow -i\partial/\partial x_a\,$.

We finally remind that the global parity
transformation is given by the product of orbital and internal parity
transformations. In our picture the internal parity 
is given by  
\begin{equation}
\breve\gamma\otimes\breve\gamma=
\left(\,
\begin{matrix}
{\bf I}_8 & 0 \spazio{0.8}\cr
0 & - {\bf I}_8 
\end{matrix}
\,\right)\,.
\label{Parity}
\end{equation}
\medskip
 
It is straightforward to verify that the global angular momentum
${\bm J}={\bm L}+{\bm S}$ and the parity 
are conserved.
Together with $\lambda$ they provide a classification of the states of
the global symmetry. Concerning the problem of canonical realization,
Poincar\'e representations and invariant scalar products we refer to
the paper \cite{LoLu}. 
It is also straightforward to observe that the conservations 
properties keep holding in the presence of interactions depending 
upon the Lorentz scalars $r_a r_a$, $q_a q_a$ and $q_a r_a$: 
this fact will be used in the next section to determine a reduced set 
of differential equations and to state the boundary value problem
in the space of the square-integrable functions on the relative 
${\bf{R^3}}$.


\sect{The spectral problem for an interacting system}
\label{Sec_spectrum}


The construction of the states with  assigned
angular momentum $(j,m)$ and given parity $(-)^j$ or $(-)^{j+1}$ 
is done as usual, by multiplying  the contributions coming from
the composition of orbital and intrinsic angular momenta
by functions of $r$. 
The expressions of the states $\Psi_{+}$ and $\Psi_{-}$
are reported in equations (\ref{StatoPari}) and (\ref{StatoDispari}) 
given in Appendix A.

The eigenvalue problem comes out when we try to determine the eight
unknown function 
\begin{eqnarray}
a_i(r)\,,~ b_i(r)\,,~ c_i(r)\,,~d_i(r)\,,~~~ 
(\,i=0,1\,) 
\label{var01}
\end{eqnarray}
in the state $\Psi$ by solving the homogeneous equation 
\begin{eqnarray}
\Bigl(\,H_0-h(r)\,\Bigr)\,\Psi=0\,,
\label{eigeneq}
\end{eqnarray}
with $h(r)$ as in (\ref{h(r)}).  
Since the parity and the angular momentum are
conserved, equation (\ref{eigeneq}) produces
two different spectral problems to be  discussed separately.

\subsection{ The spectral problem for the state $\Psi_{+}\,$}
\label{evenPsi}

When
substituting  $\Psi=\Psi_{+}$ in (\ref{eigeneq}), 
we obtain a system of equations given by the vanishing of
the coefficients of 
the different spherical harmonics in each component of the resulting
vector. This system is composed of thirty-four first order 
differential equations, but, as one should expect,
only eight of them are
independent.
The detailed expressions for the eight equations are written in 
Appendix B. 

It turns out that appropriate changes of the initial variables
can help in giving the system a much simpler and readable form. To this
purpose, we first define the sums and differences
\begin{eqnarray}
s_\pm(r)=s_0(r)\pm s_1(r)\,~~~~~(\,s=a,b,c,d\,)
\label{variabilipm}
\end{eqnarray}
and we consider the following linear combinations of the $c_\pm(r)$ and 
$d_\pm(r)$: 
\begin{eqnarray}
&{}& {u_+}(r)=-\,{\displaystyle \frac {\sqrt{j}\,{c_+}(r) - 
\sqrt{j + 1}\,{d_+}(r)}{\sqrt{1 + 2\,j}}}\spazio{1.2}\cr
 &{}& {u_-}(r)=-\,{\displaystyle \frac {\sqrt{j}\,{c_-}(r) - 
\sqrt{j + 1}\,{d_-}(r)}{\sqrt{1 + 2\,j}}}\spazio{1.2}\cr
&{}& {v_+}(r)=-\,{\displaystyle \frac {\sqrt{j + 1}\,{c_+}(r)
 + \sqrt{j}\,{d_+}(r)}{\sqrt{1 + 2\,j}}}\spazio{1.2}\cr
 &{}& {v_-}(r)=-\,{\displaystyle \frac {\sqrt{j + 1}\,{c_-}(r)
 + \sqrt{j}\,{d_-}(r)}{\sqrt{1 + 2\,j}}}\,.
 \label{varuvpm}
\end{eqnarray}
A rather simple formulation for the system can then be given in 
terms of $a_\pm(r)$, $b_\pm(r)$, $u_\pm(r)$ and $v_\pm(r)$.
Indeed, from a straightforward computation, we find that
the eight independent equations split into four 
algebraic relations and four  first order differential equations.
The former read
\begin{eqnarray}
&{}& {\sqrt{j(j + 1)}\,{a_+}(r)}
  - {\displaystyle \frac {r}{2}}\,({h}(r) \,{v_+}(r)+\mu \,{v_-}(r))
  =0\spazio{1.}\cr
&{}&   {\sqrt{j(j + 1)}\,{b_-}(r)}  + {\displaystyle \frac {r}{2}}\,
(\mu\,{u_+}(r)  + {h}(r) \,{u_-}(r))
=0 \spazio{1.}\cr
&{}& M\,{a_+}(r) - {h}(r)\,{a_-}(r)=0\spazio{1.}\cr
&{}& {h}(r)\,{b_+}(r) - M\,{b_-}(r)=0
\label{relalgeven}
\end{eqnarray}
while the differential equations are 

\begin{eqnarray}
&{}&{\displaystyle\frac {d{a_+}(r)}{dr}} + {\displaystyle \frac {
\sqrt{j(j + 1)}\,\mu }{r\,{h}(r)}}\,{b_-}(r)  
+ 
\,{\displaystyle \frac {( - {
h}(r)^{2} + \mu ^{2})}{2\,{h}(r)}}\,{u_+}(r) =0
\spazio{1.2}\cr
&{}&\sqrt{j}\,\Bigl(\,{\displaystyle\frac {d{b_-}(r)}{dr}} - 
{\displaystyle 
\frac {j{b_-}(r)}{r}}\Bigr)  - {\displaystyle \frac {\sqrt{j + 1}
}{2}}\, \Bigl(\mu {u_+}(r)+ {h}(r){u_-
}(r)\,\Bigr)
 + {\displaystyle 
\frac {\sqrt{j}}{2}}\,\Bigl( \,\mu \,{v_+}(r) +{h}(r)\,{v_-}(r)\,
\Bigr) 
\mbox{} =0\spazio{1.2}\cr
&{}&{\displaystyle\frac {d{u_+}(r)}{dr}} - {\displaystyle 
\frac {M - {h}(r)}{2}} 
\,\Bigl(\,{a_+}(r)+{a_-}(r)\,\Bigr)
  + {\displaystyle 
\frac {2}{r}}\,{u_+}(r)
- {\displaystyle \frac {j\,\sqrt{j + 
1}}{r}}\,{v_+}(r) =0\spazio{1.2}\cr
&{}&{\displaystyle\frac {d{v_-}(r)}{dr}} + {\displaystyle 
\frac {M - {h}(r)}{2}} 
\,\Bigl(\,{b_+}(r)+{b_-}(r)\,\Bigr)
 - {\displaystyle \frac {
\sqrt{j(j + 1)}}{r}}\,{u_-}(r)+ {\displaystyle 
\frac {1}{r}}\,{v_-}(r) =0
\label{diffeqeven}
\end{eqnarray}

By means of (\ref{relalgeven}) we  eliminate the four
variables that are not differentiated, obtaining  a system of four
first order differential equations in the four unknown functions
$a_+(r)$, $b_-(r)$, $u_+(r)$ and $v_-(r)$. 
If we arrange the latter respectively as the four components of a vector
$Y(r)\equiv{}^t\!(y_1(r),y_2(r),y_3(r),y_4(r))$, the system is given the 
compact form
\begin{eqnarray}
{\displaystyle \frac {dY(r)}{dr}}+{\cal M}\,Y(r)=0\,,
\label{sistema_mat}
\end{eqnarray}
where ${\cal M}$ is a matrix with general structure 
\begin{eqnarray}
{\cal M}= \left[ 
{\begin{array}{cccc}
\phantom{-}0~ & \phantom{-}E(r)~ & \phantom{-}F(r)~ & \phantom{-}0~ 
\spazio{1.2}\\
\phantom{-}E(r)~ & \phantom{-}{\displaystyle \frac {1}{r}}~  & 
\phantom{-}0~ &  - F(r)~ \spazio{1.2}\\ [2ex]
\phantom{-}{G}(r)~ & \phantom{-}0~ & \phantom{-}
{\displaystyle \frac {2}{r}}~  & \phantom{-}E(r)~ \spazio{1.2}\\ [2ex]
\phantom{-}0~ &  - {G}(r)~ & \phantom{-}E(r)~ & 
\phantom{-}{\displaystyle \frac {1}{r}}~ 
\end{array}}
 \right] 
\label{sysmat}
\end{eqnarray}
The explicit expressions of $E(r)$, 
$F(r)$ and $G(r)$ for the even case  
are the following ones
\begin{subequations}
\label{equaUVW}
\begin{eqnarray}
&{}& E_{\mathrm e}(r)={\displaystyle 
\frac {\sqrt{j(j + 1)}\,\mu }{r\,{h}(r)}}
\spazio{1.}
\label{equa_U}\\
&{}& F_{\mathrm e}(r)= - {\displaystyle \frac {h(r)}{2}}\,\Bigl(1 - 
{\displaystyle \frac {\mu ^{2}}{{h}^2(r) }} \Bigr)
\spazio{1.}
\label{equaV}\\
&{}& G_{\mathrm e}(r)={\displaystyle \frac {h(r)}{2}}\,\Bigl(1 - 
{\displaystyle 
\frac {r^{2}\,M^{2} + 4\,j\,(j + 1)}{r^{2}\,
{h}^2(r) }}\,\Bigr)\,.
\label{equaW}
\end{eqnarray}
\end{subequations}
\smallskip
Specifying the matrix elements (\ref{equaUVW}) 
to the case of the Coulomb interaction, where
\begin{eqnarray}
{h}(r)=\lambda+\alpha /r\,,
\label{Coulomb}
\end{eqnarray}
and where $\alpha$ is the fine structure constant, we have
\begin{subequations}
\label{equaUVW_C}
\begin{eqnarray}
&{}& E_{{\mathrm e},\mathrm {Coul}}(r)={\displaystyle 
\frac {\sqrt{j(j + 1)}\,\mu }
{\lambda\,r+\alpha}}
\spazio{1.}\\
\label{equaU_C}
&{}& F_{{\mathrm e},\mathrm {Coul}}(r)= - {\displaystyle \frac
{(\lambda r+\alpha)^2-\mu^2r^2}
{2r\,(\lambda r+\alpha)}} 
\spazio{1.}\\
\label{equaV_C}
&{}& G_{{\mathrm e},\mathrm {Coul}}(r)=  
{\displaystyle \frac {(\lambda r+\alpha)^2-M^2r^2-4j(j+1)} 
{2r\,(\lambda r+\alpha)}~~~~~~~   }
\label{equaW_C}
\end{eqnarray}
\end{subequations}

Some remarks are in order. 
First of all, from (\ref{sysmat}) and (\ref{equaUVW}) we see that the 
transformation
\begin{eqnarray}
\mu\rightarrow -\mu\,,\qquad y_1(r)\rightarrow -y_1(r)\,,\qquad
y_3(r)\rightarrow -y_3(r)
\end{eqnarray}
\noindent
is a symmetry of the system, corresponding to the fact that the
ordering of the two particle is irrelevant.

In the second place, from 
(\ref{equa_U}), we see that $E_{\mathrm e}(r)=0$ 
either for $j=0$ or $\mu=0$. In these cases 
the fourth order system splits 
into separate second order subsystems, one for the functions 
$y_1(r)$ and $y_3(r)$, the other for $y_2(r)$ and $y_4(r)$. The
subsystems, in turn, can be presented as the following two independent 
second order differential equations
\begin{subequations}
\label{U0pari}
\begin{eqnarray}
&{}& {\displaystyle\frac {d^{2}}{dr^{2}}}\,{y_{1}}(r) 
 +\Bigl({\displaystyle \frac 2r}-{\displaystyle \frac {\displaystyle
 \frac {d}{dr}\,F(r)}{F(r)}}
 \Bigr)\,  
 {{\displaystyle\frac {d}{dr}}\,{y_{
1}}(r)}
-{F}(r)
\,{G}(r)\,{y_{1}}(r)=0~~~~~~~\spazio{1.0}
\label{U0pari_1}\\
&{}& {\displaystyle\frac {d^{2}}{dx^{2}}}\,{y_{2}}(x)  
 +\Bigl({\displaystyle \frac 2r}-
 {\displaystyle \frac {\displaystyle\frac {d}{dr}\,F(r)}{F(r)}}
 \Bigr)\,  
 {{\displaystyle\frac {d}{dr}}\,{y_{2}}(r)}
- \Bigl(\,
{\displaystyle \frac {\displaystyle\frac {d}{dr}\,F(r)}{r\,F(r)}}
 + {F}(r)\,{
G}(r) \,\Bigr  )\,{y_{2}}(r)=0
\label{U0pari_2}
\end{eqnarray}
\end{subequations}
\smallskip
\noindent
As discussed in Section \ref{Sec_dirac}, the Dirac limit 
shows that
for $j=0$ only (\ref{U0pari_1}) makes sense. 

Although the linear system (\ref{sistema_mat}) and, in particular,
the equation (\ref{U0pari_1}) do not look extremely
complicated, they still are difficult enough to prevent the
possibility of an easy analytical solution. 
For instance, according to a somewhat old fashioned -- but 
actually still the most up to date -- classification of differential
equations according to the number of their elementary singularities
(for definitions and detail we refer to the classical 
text \cite{Ince}), equations (\ref{U0pari}) with the Coulomb
coefficients (\ref{equaUVW_C}) present eight
elementary singularities, while analytical properties of the solutions
have been studied and classified for differential equations having
at most six elementary singularities. 
Numerical techniques are therefore necessary
if we want to determine the spectral properties, 
both for equal and different masses.

\subsection { The spectral problem for the state $\Psi_{-}\,$}

The differential equations generated by the action of the Hamiltonian 
on the odd state are easily derived by observing
that the transformation (\ref{StatoDispari})  amounts to changing 
$m_1$ into
$-m_1$, which means $\mu\rightarrow -M$ and $M\rightarrow -\mu$. 
The general structure for the matrix of the odd linear system,
therefore, remains the one given in (\ref{sysmat}). The matrix elements,
on the other hand, are as follows:
\begin{subequations}
\label{equaUVW_odd}
\begin{eqnarray}
&{}& E_{\mathrm o}(r)=-{\displaystyle \frac {\sqrt{j(j + 1)}\,M }{r\,{h}(r)}}
\spazio{1.}
\label{equaU_odd}\\
&{}& F_{\mathrm o}(r)= - {\displaystyle \frac {h(r)}{2}}\,\Bigl(1 - 
{\displaystyle \frac {M ^{2}}{{h}^2(r) }} \Bigr)
\spazio{1.}
\label{equaV_odd}\\
&{}& G_{\mathrm o}(r)={\displaystyle \frac {h(r)}{2}}\,\Bigl(1 - 
{\displaystyle 
\frac {r^{2}\,\mu^{2} + 4\,j\,(j + 1)}{r^{2}\,
{h}^2(r) }}\,\Bigr)\,.
\label{equaW_odd}
\end{eqnarray}
\end{subequations}
and also the Coulomb explicit expressions are obtained from 
(\ref{equaUVW_C}) by the exchanges $(M,\mu)\rightarrow(-\mu,-M)$.
The system decouples for  $j=0$, but, 
in contrast to what occurs in the even case,
this circumstance is not realized
when the particles have equal masses.


\sect{The Dirac limit}
\label{Sec_dirac}


In this section we prove that (\ref{sistema_mat}-\ref{sysmat}) 
has the correct
Dirac limit. This means that when the mass of one 
of the two interacting particles, say $m_1$, 
tends to infinity, the system
reduces to the Dirac equation for a charged fermion in an
external potential. 

The limit is defined by assuming 
that the mass $m_2\equiv m_e$ remains finite
and gives the mass scale. The 
rescaling of the mass parameters of the system is then 
\begin{eqnarray}
\mu=\displaystyle\frac{2\rho\,m_e}{1-\rho}\,,\qquad
M=\displaystyle\frac{2m_e}{1-\rho}\,,
\label{massrescale}
\end{eqnarray}
We also take a dimensionless independent variable 
\begin{eqnarray}
x=m_e\,r
\label{variabilex}
\end{eqnarray}
and define a coefficient $\eta$ by means of the relation
\begin{eqnarray}
h(x)=\Bigl(\eta -v(x)
+\displaystyle\frac{1+\rho}{1-\rho}\Bigr)\,m_e\,
\label{h_x}
\end{eqnarray}
where $v(x)$ is the potential expressed in terms of $x$ and
divided by $m_e$. It proves it convenient to define the shorthand
\begin{eqnarray}
\eta(x)=\eta-v(x)\,.
\label{eta}
\end{eqnarray}
We now take the limit $\rho\rightarrow 1$ on the matrix elements. For 
the even system in the dimensionless variable $x$, we find 
\begin{eqnarray}
&{}& E_{\mathrm e}(x)\,\rightarrow ~\sqrt{j(j+1)}/x\spazio{0.8}\cr
&{}& F_{\mathrm e}(x)\,\rightarrow ~-(1+\eta(x))\spazio{0.8}\cr
&{}& G_{\mathrm e}(x)\,\rightarrow ~\eta(x)-1\,.
\label{DiarcPari}
\end{eqnarray}
For the odd system:
\begin{eqnarray}
&{}& E_{\mathrm o}(r)\,\rightarrow ~-\sqrt{j(j+1)}/x\spazio{0.8}\cr
&{}& F_{\mathrm o}(r)\,\rightarrow ~(1-\eta(x))\spazio{0.8}\cr
&{}& G_{\mathrm o}(r)\,\rightarrow ~(1+\eta(x))\,.
\label{DiarcDispari}
\end{eqnarray}
Let us prove that the equations obtained by the use of 
(\ref{DiarcPari}) 
and (\ref{DiarcDispari}) are actually equivalent to pairs of
Dirac equations \cite{LL}, that we write in our rescaled variables
as
\begin{equation}
{\displaystyle\frac{d}{dx}}\,
\left({\begin{array}{c} x \,f(x) \spazio{1.0}\\ x \,g(x)
\end{array}}\right)
+
\left({\begin{array}{cc} {\displaystyle\frac{\kappa}{x}} 
& -(1+\eta-v(x)) \spazio{1.0}\\
 (-1+\eta-v(x)) & {-\displaystyle\frac{\kappa}{x}}
 \end{array}}\right)\,
 \left({\begin{array}{c} x \,f(x) \spazio{1.0}\\ x \,g(x)
\end{array}}\right)
\label{dirac_sys}
\end{equation}
\noindent It is not difficult to realize that 
a mixing is necessary in order to decouple the fourth order 
system in two second order subsystems to be compared to the Dirac
equation.
This is physically clear, since
from a description focusing on the angular 
momentum properties of the
bosonic global system, we want to determine the equation for a
fermionic object that constitutes only a part of the compound.
The required linear transformation is generated by the 
orthogonal constant matrix 

\begin{eqnarray}
 {\cal U}={\displaystyle \frac 1{\sqrt{2\,j+1}}}\,\left( 
{\begin{array}{cccc}
0 & 0 & \sqrt{j} & \sqrt{j+1} \spazio{1.}\cr
0 & 0 & \sqrt{j+1} & -\sqrt{j}\spazio{1.}\cr 
\sqrt{j+1} & -\sqrt{j} & 0 & 0\spazio{1.}\cr 
\sqrt{j} & \sqrt{j+1} & 0 & 0 
\end{array}}
 \right)\spazio{1.2} 
\label{diagodirac}
\end{eqnarray}
\noindent By defining 
\begin{equation}
z_i(x)=\sum\limits_{j=1}^4~({\cal U}^{-1})_{ij}\,\,
x\,y_j(x)
\end{equation}
it is straightforward to see that both in the even 
and the odd case the
equations for the variables ${z_{1}}(x)$ and ${z_{4}}(x)$ decouple 
from the equations for ${z_{2}}(x)$ and ${z_{3}}(x)$. 
The comparison
with (\ref{dirac_sys}) shows that, 
the subsystems obtained in the even case are respectively the
Dirac equations with $\kappa=-(j+1)$ and $\kappa=j$ 
(here we must change ${z_{3}}(x)$ into $-{z_{3}}(x)$):
these are the two allowed values of  $\kappa$ 
for the states whose orbital momentum is 
$j$ in the Dirac spectroscopy,  \cite{LL}. 
In the odd case  the decoupled systems agree respectively with the Dirac
equations with $\kappa=j+1$ (here changing ${z_{4}}(x)$ into
$-{z_{4}}(x)$) and $\kappa=-j$: these values correspond to
orbital angular momenta equal to $j+1$ and $j-1$.

The previous discussion clearly shows that the complete treatment is
continuous in the parameter $1-\rho\,$: indeed, when this 
parameter is small, it can be effectively 
used computing perturbative corrections to the
Dirac equation. Thus, although the general features of the spectrum -- 
{\it {e.g.}} its boundedness from below -- are clearly dependent upon 
the potential function $v(x)$ and can be studied in each particular case 
by well established criteria \cite{DS}, the continuity in $1-\rho\,$ of
the differential operator implies the continuity in $1-\rho\,$ of its
spectrum that therefore shares the qualitative properties of the
limiting case.
We report in Appendix C 
the calculation of the first
order correction to the Hydrogen ground state
and we find an excellent agreement with
the numerical results.


\sect{Discussion of the results}
\label{Sec_results}


In this section we present the results obtained. We 
display the tables containing the values of the variable $\rho$ and
the corresponding level with ten meaningful figures; the 
corresponding plots are then drawn. 
The energy parameter $w$ that appears both in the tables and in the
plots is related to the parameter $\eta$ defined in (\ref{eta}) by the
relation
\begin{eqnarray}
\eta=1+{\displaystyle\frac 12}\,(1+\rho)\,\alpha^2\,w
\label{etaw}
\end{eqnarray}
\noindent
In terms of the eigenvalue $\lambda$, this equation 
 is equivalent to $\lambda=M+m_R\,\alpha^2\,w$, 
%
%
%
%
%
where $m_R$ is the classical reduced mass $m_R=m_1\,m_2/(m_1+m_2)\,$. 
This means that the correction of energy levels due to the classical
reduction to the center of mass frame is already accounted by the scale.
Hence, the dependence upon $\rho$ shown below is a genuine relativistic
effect.
  
The results will be given in atomic units, {\it i.e.} $m_e=1$ in
addition to $\hbar=1$ and $c=1$.
The values of $\rho$ with greater physical relevance
are  $\rho=\rho_e$, corresponding to the hydrogen atom,
$\rho=\rho_\mu$, that refers to the mesic atom and $\rho=0$,
for the positronium. The value of $w$ 
for $\rho=1$ has been taken
from the analytical formula for the Dirac spectrum.
It is natural to begin from the ground state. We find that for
any value of the two fermion masses the lowest energy level is
degenerate with multiplicity two and the corresponding states 
are the first ({\it {i.e.}} with lowest eigenvalue) even state with 
$j=0$ and the first odd state with $j=1$. 
We specify the results in the Table here below and we shall comment on
the degeneracy later on. 
\medskip\medskip

\begin{center}
\begin{tabular}[t]{l|l}
$\rho$\phantom{x}  &\phantom{xxi}  
$w_{\,+,\,\,j\,=\,0,\,\,{\mathrm I}}$ 
\spazio{0.6}\\  \hline
\phantom{}  &\phantom{}\\
0.0\phantom{x}  &\phantom{x}   -.4999950109    \\
0.2\phantom{x}  &\phantom{x}   -.4999954766      \\
0.4\phantom{x}  &\phantom{x}  -.4999968737        \\
0.6\phantom{x}  &\phantom{x}  -.4999992025           \\
$\rho_\mu$\phantom{x}  &\phantom{x}   -.5000024182         \\
$\rho_e$\phantom{x}  &\phantom{x}  -.5000066312               \\
1.0\phantom{x}  &\phantom{x}  -.5000066566 
\end{tabular}
\qquad\qquad\qquad
\begin{tabular}[t]{l|l}
$\rho$\phantom{x}  &\phantom{xxi}  
$w_{\,-,\,\,j=\,1\,,\,\,{\mathrm {I}}}$  \spazio{0.6}\\  \hline
\phantom{}  &\phantom{}\\
0.0\phantom{x}  &\phantom{x}   -.4999950109    \\
0.2\phantom{x}  &\phantom{x}   -.4999954766      \\
0.4\phantom{x}  &\phantom{x}  -.4999968737        \\
0.6\phantom{x}  &\phantom{x}  -.4999992025           \\
$\rho_\mu$\phantom{x}  &\phantom{x}   -.5000024182         \\
$\rho_e$\phantom{x}  &\phantom{x}  -.5000066312               \\
1.0\phantom{x}  &\phantom{x}  -.5000066566 
\label{spettroj1oddFund}
\end{tabular}
\end{center}
\medskip\medskip

\noindent 
Here $\rho_e=0.998911$ corresponds to the hydrogen atom data and 
$\rho_\mu=0.797576$ refers to the mesic atom.
The plot of the ground state {\it vs.} $\rho$ is shown in Fig.1. 
The difference $w_{\,+,\,j\,=\,0,\,\,{\mathrm {I}}\,}(\rho_e)-
w_{\,+,\,j\,=\,0,\,{\mathrm {I}}\,}(1)=0.254\,\cdot\,10^{-7}$ 
should be compared
with the value $0.253\,\cdot\,10^{-7}$ obtained in Appendix C by a
perturbative calculation.

\medskip\medskip\medskip\medskip\medskip
\begin{center}
\includegraphics[height=6.cm,width=7cm]{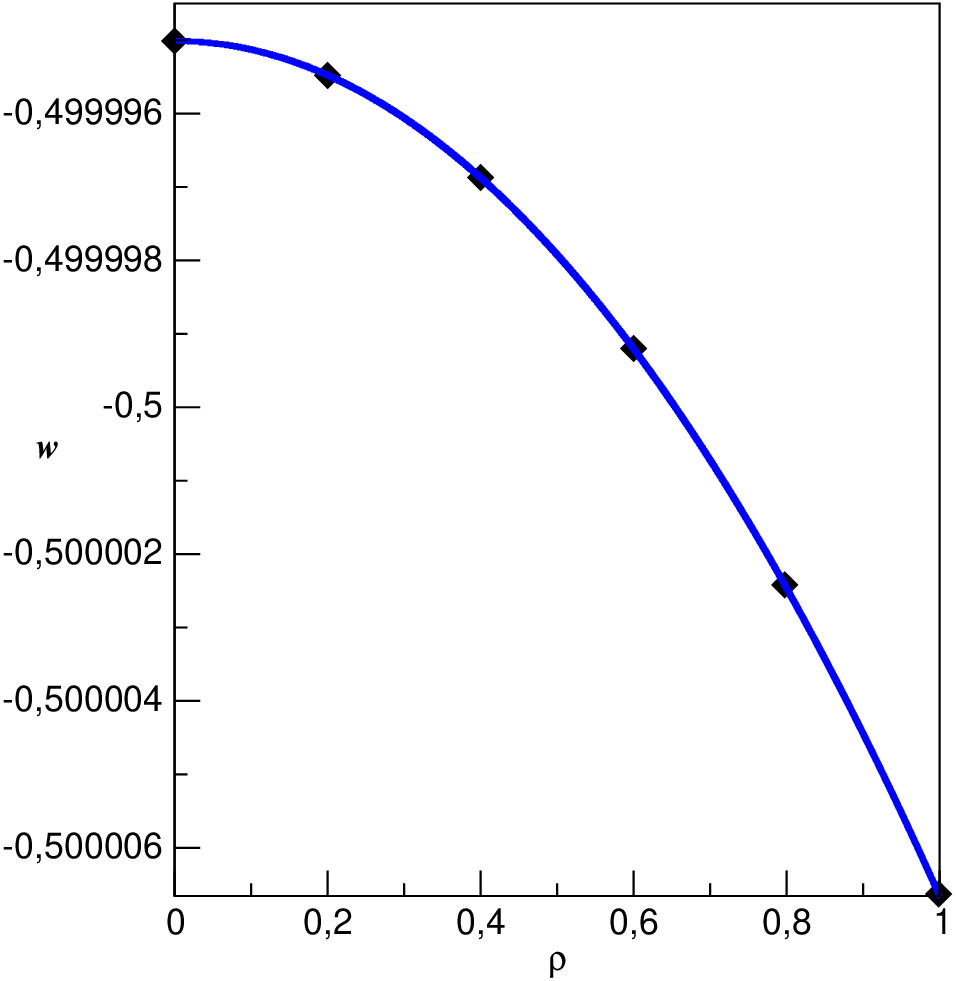}
\end{center}

\begin{footnotesize}
\noindent { 
FIG. 1 The dependence of the degenerate ground state upon  $\rho\,$:
the levels $w_{\,+,\,j\,=\,0,\,\,{\mathrm {I}}}$  and 
$w_{\,-,\,j\,=\,1,\,\,{\mathrm {I}}}\,$.}
\end{footnotesize}

\medskip\medskip\medskip

The triplet given by the first even excited state with $j=0$ --  
corresponding to the state
$2s_{1/2}$ in the Dirac limit -- together
with the even states that in the Dirac limit reduce to $2p_{1/2}$ and
$2p_{3/2}$ is presented in the first plot of Fig.2.  
Their numerical values are 
respectively reported in the Table that follows:
\medskip
\begin{center}
\begin{tabular}[t]{l|l}
$\rho$\phantom{x}  &\phantom{xxi}  
$w_{\,+,\,\,j\,=\,0,\,\,{\mathrm {II}}}$  \spazio{0.6}\\  \hline
\phantom{}  &\phantom{}\\
0.0\phantom{x}  &\phantom{x}   -.1249996884     \\
0.2\phantom{x}  &\phantom{x}  -.1249997840       \\
0.4\phantom{x}  &\phantom{x}  -.1250000710        \\
0.6\phantom{x}  &\phantom{x}   -.1250005493            \\
$\rho_\mu$\phantom{x}  &\phantom{x} -.1250012098           \\
$\rho_e$\phantom{x}  &\phantom{x}  -.1250020750                 \\
1.0\phantom{x}  &\phantom{x}   -.1250020802          
\label{spettroj0}
\end{tabular}
\qquad\qquad
\begin{tabular}[t]{l|l}
$\rho$\phantom{x}   &\phantom{xxi}  
$w_{\,+,\,\,j\,=\,1,\,\,{\mathrm I}}$  \spazio{0.6}\\  \hline
\phantom{}  &\phantom{}\\
0.0\phantom{x}  &\phantom{x}   -.1250005200     \\
0.2\phantom{x}  &\phantom{x}   -.1250006292      \\
0.4\phantom{x}  &\phantom{x}   -.1250008727       \\
0.6\phantom{x}  &\phantom{x}  -.1250012005           \\
$\rho_\mu$\phantom{x}  &\phantom{x}    -.1250015984         \\
$\rho_e$\phantom{x}  &\phantom{x}    -.1250020774              \\
1.0\phantom{x}  &\phantom{x} -.1250020802          
\end{tabular}
\qquad\qquad
\begin{tabular}[t]{l|l}
$\rho$\phantom{x}   &\phantom{xxi}  
$w_{\,+,\,\,j\,=\,1,\,\,{\mathrm {II}}}$  \spazio{0.6}\\  \hline
\phantom{}  &\phantom{}\\
0.0\phantom{x}  &\phantom{x}   -.1250002427      \\
0.2\phantom{x}  &\phantom{x}   -.1250002030        \\
0.4\phantom{x}  &\phantom{x}  -.1250001675          \\
0.6\phantom{x}  &\phantom{x}   -.1250001864            \\
$\rho_\mu$\phantom{x}  &\phantom{x}  -.1250002673           \\
$\rho_e$\phantom{x}  &\phantom{x}  -.1250004151                   \\
1.0\phantom{x}  &\phantom{x}    -.1250004160          
\label{spettroj1}
\end{tabular}
\end{center}

\medskip\medskip\medskip
\begin{center}
\includegraphics[height=6.cm,width=7cm]{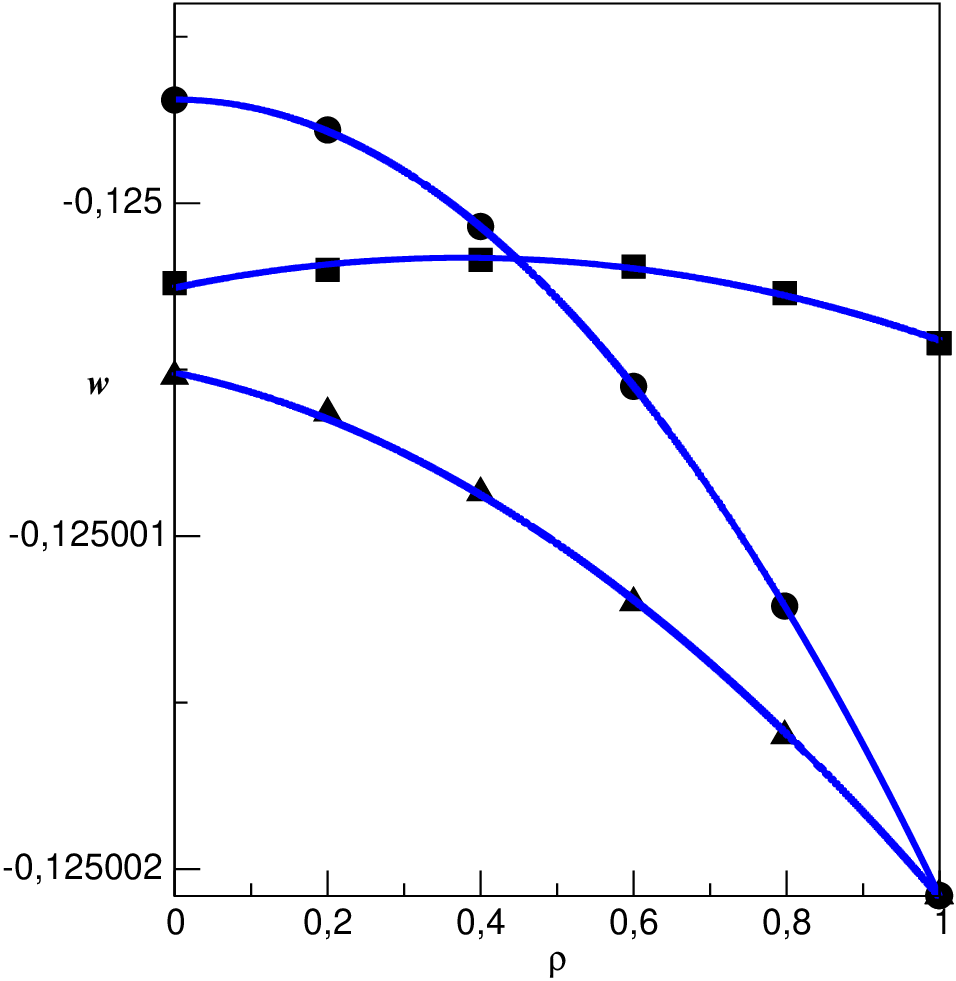}
$~~~$\includegraphics[height=6.cm,width=7cm]{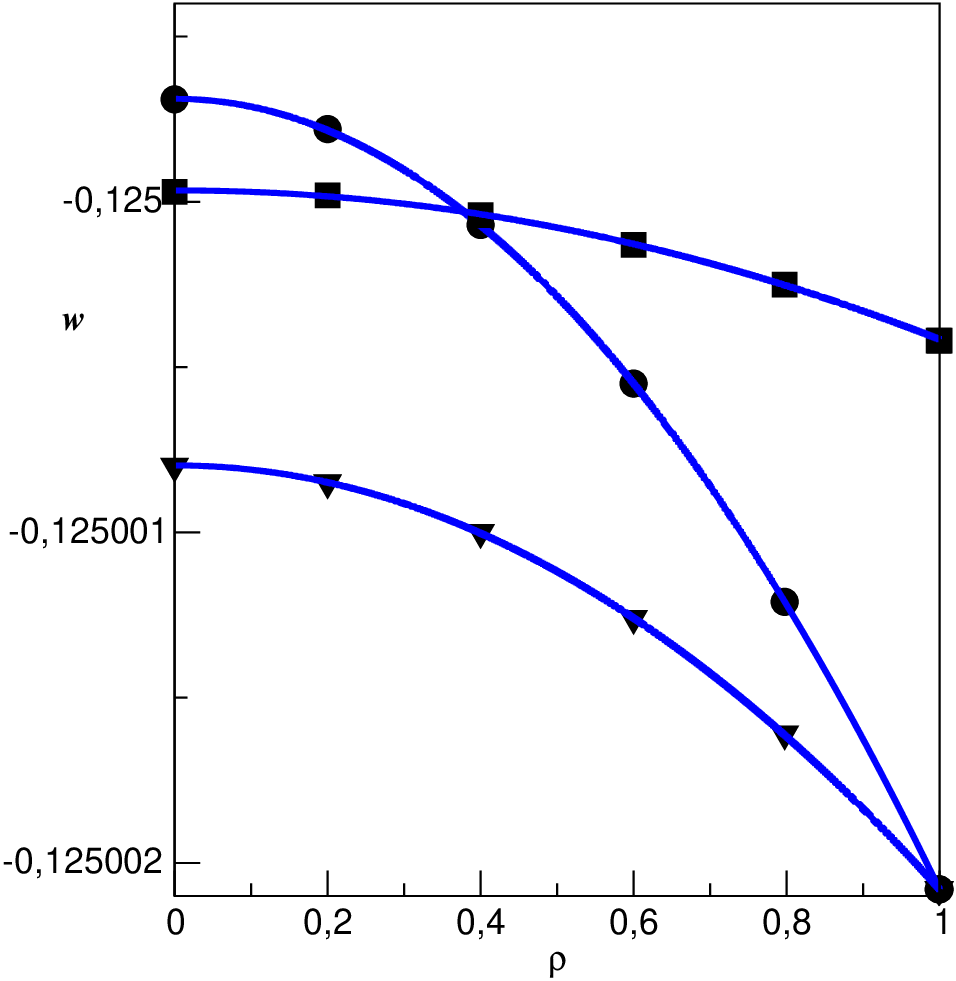}$~~~$
\end{center}

\medskip

\begin{footnotesize}
\noindent
{FIG. 2 The spectral curves for varying masses from the even and odd 
states. In the left plot (even states), 
circles 
\noindent\phantom{FIG. 2} represent the curve 
$\,~w_{\,+,\,j\,=\,0,\,\,{\mathrm {II}}}~\,$
that reduces to the state $\,~2s_{1/2}~\,$ in the
Dirac limit, triangles the curve 

\noindent\phantom{FIG. 2} $w_{\,+,\,j\,=\,1,\,\,{\mathrm {I}}}$
reducing to the state $2p_{1/2}$ and 
squares the curve $w_{\,+,\,j\,=\,1,\,\,{\mathrm {II}}}$ 
reducing to ${2p_{3/2}}$. In the right 
\noindent\phantom{FIG. 2} plot (odd states),
circles to refer to  $w_{\,-,\,j\,=\,1,\,\,{\mathrm {I}}}\,$,    
triangles to $w_{\,-,\,j\,=\,0,\,\,{\mathrm {I}}}\,$, squares to 
$w_{\,-,\,j\,=\,2,\,\,{\mathrm {I}}}\,$.}

\end{footnotesize}

\medskip\medskip

In the final Table we give the values for the three states that 
constitute the odd counterpart of the even triplet and that are shown in 
the second plot of Fig.2. They
arise from the three different values $j=0,1,2\,$ of the angular
momentum. 
\medskip
\begin{center}
\begin{tabular}[t]{l|l}
$\rho$\phantom{x}  &\phantom{xxi}  
$w_{\,-,\,\,j\,=\,0,\,\,{\mathrm {I}}}$  \spazio{0.6}\\  \hline
\phantom{}  &\phantom{}\\
0.0\phantom{x}  &\phantom{x}   -.1250007974    \\
0.2\phantom{x}  &\phantom{x}  -.1250008487       \\
0.4\phantom{x}  &\phantom{x}    -.1250010026       \\
0.6\phantom{x}  &\phantom{x}  -.1250012592           \\
$\rho_\mu$\phantom{x}  &\phantom{x} -.1250016134           \\
$\rho_e$\phantom{x}  &\phantom{x} -.1250020774                 \\
1.0\phantom{x}  &\phantom{x}   -.1250020802         
\label{spettroj2odd}
\end{tabular}
\qquad\qquad
\begin{tabular}[t]{l|l}
$\rho$\phantom{x}   &\phantom{xxi}  
$w_{\,-,\,\,j\,=\,1,\,\,{\mathrm {II}}}$  \spazio{0.6}\\  \hline
\phantom{}  &\phantom{}\\
0.0\phantom{x}  &\phantom{x} -.1249996884       \\
0.2\phantom{x}  &\phantom{x}  -.1249997840        \\
0.4\phantom{x}  &\phantom{x}   -.1250000710       \\
0.6\phantom{x}  &\phantom{x}    -.1250005493           \\
$\rho_\mu$\phantom{x}  &\phantom{x}  -.1250012098          \\
$\rho_e$\phantom{x}  &\phantom{x} -.1250020750                  \\
1.0\phantom{x}  &\phantom{x}     -.1250020802         
\end{tabular}
\qquad\qquad
\begin{tabular}[t]{l|l}
$\rho$\phantom{x}   &\phantom{xxi}  
$w_{\,-,\,\,j\,=\,2,\,\,{\mathrm {I}}}$  \spazio{0.6}\\  \hline
\phantom{}  &\phantom{}\\
0.0\phantom{x}  &\phantom{x}   -.1249999654      \\
0.2\phantom{x}  &\phantom{x}   -.1249999834        \\
0.4\phantom{x}  &\phantom{x}   -.1250000375          \\
0.6\phantom{x}  &\phantom{x}   -.1250001276            \\
$\rho_\mu$\phantom{x}  &\phantom{x}  -.1250002520           \\
$\rho_e$\phantom{x}  &\phantom{x}  -.1250004150                   \\
1.0\phantom{x}  &\phantom{x}    -.1250004160          
\label{spettroj1odd}
\end{tabular}
\end{center}
\medskip\medskip

From the data we discover one more degeneracy: indeed the even state
$w_{\,+,\,j\,=\,0,\,\,{\mathrm {II}}}$ and the odd state
$w_{\,-,\,j\,=\,1,\,\,{\mathrm {II}}}$, obviously equal in the
Dirac limit, remain degenerate for any value of $\rho\,$. 
This occurrence, in the present case and for the fundamental state 
as well, is certainly not evident from a mathematical or computational 
point of view: indeed the equations that have been 
solved to find the even spectral curves and the systems that give the 
the odd ones are completely unrelated.
We can however understand this apparently peculiar behavior by thinking 
of the non relativistic limit. Here, unless the spin-orbit coupling is 
explicitly included, the total spin and the orbital angular momentum are
separately conserved. Moreover, if no spin-spin interaction is present,
the relative orientations of the two spins between themselves and with
respect to the orbital angular momentum are effective in determining 
the parity, but have no influence on energy of the state. The result is
that the even and the odd states are degenerate with respect to parity.
This degeneracy survives in the Dirac limit:
although the spin-orbit interaction is now directly accounted for, 
one of the two constituent particles is infinitely heavy 
and thus at rest, so that only the spin of the light particle matters. 
The situation is different in our relativistic equation,
when both constituents have finite mass and 
each of them gives its own contribution to the spin-orbit
coupling: in fact we obtain different shifts for the even and 
odd terms, but for those states that in the non relativistic limit have 
a vanishing orbital
angular momentum. In the particular case of the Positronium,     
we see that the two states $w_{\,+,\,j\,=\,0,\,\,{\mathrm {I}}}$  and 
$w_{\,-,\,j\,=\,1,\,\,{\mathrm {I}}}$ correspond to Parapositronium and 
Orthopositronium respectively. Their degeneracy could be removed by
introducing the spin-spin interaction, responsible for the hyperfine
splitting. We shall comment on this subject in Appendix E. 

Coming back to the figures, we observe that the solid curves are 
the parabolas obtained by the best fit of the data and they
are in excellent agreement with the data themselves. 
The perturbative formula given in \cite{Chi}
for the Coulomb Lamb shift, $\delta w=(\alpha^4/12)(1-\rho)/(1+\rho)^2$, 
can reproduce a parabolic behavior only for
$\rho$ in the neighborhood of zero or unity. Only the second of these 
possibilities is actually realized: this shows the necessity of
a more detailed numerical analysis when the masses become closer and
closer. 
It is nevertheless interesting to observe the increasing impact of the
relativistic effects with the decreasing of the mass difference. In
particular it has to be noticed the crossing of the levels
$w_{\,+,\,j\,=\,0,\,\,{\mathrm {II}}}$ and 
$w_{\,+,\,j\,=\,1,\,\,{\mathrm {II}}}$ as well as
$w_{\,-,\,j\,=\,1,\,\,{\mathrm {II}}}$ and 
$w_{\,-,\,j\,=\,2,\,\,{\mathrm {I}}}\,$: 
this, for instance, 
reflects the necessity of the different method in
use to classify the Positronium states with respect to the Dirac scheme.    

We conclude this discussion of the results with some observations on
a possible formulation of a more realistic model involving a minimal
coupling that accounts for the complete electrodynamical interaction. 
In fact this type of coupling can be arranged
quite easily in our quantum mechanical framework and, 
if we study the case of a closed two-body system and we
do want to keep a pure relativistic quantum mechanical description, 
we can try to express the vector potential itself in terms of the 
dynamical variables of the two particles.  
The Li\'enard-Wiechert potentials provide the most natural answer to
the raised question. Unfortunately they introduce a delay in the
propagation of the interaction that induces a considerable amount
of difficulty in the solution of the corresponding wave equation and
indeed the  necessity of rendering its solution less tough
is a major source of approximations, that, in most cases, produce
not very satisfactory results (see \cite{Chi} and references therein for
a discussion on the subject). Indeed if one stops at the lowest order of
approximation in the Li\'enard-Wiechert potentials, neglecting the
 the propagation delay, one finds an interaction energy
\begin{eqnarray}
V_{\mathrm {em}}(r)=
{\displaystyle \frac\alpha r} + {\displaystyle \frac\alpha {2r}}\,
\Bigl(\,{v_{(1)}}_{a}\,{v_{(2)}}_{a}+(n_a {v_{(1)}}_{a})\,
(n_a {v_{(2)}}_{a})\,\Bigr)
\end{eqnarray}
that has to be substituted to $V(r)$ in (\ref{h(r)}).
According to the Dirac prescription, when switching to
quantum mechanics, the particle velocities $v_{(i)a}$ will be substituted by
the $\alpha_{(i)a}=-\,\breve{\gamma}_{(i)}{\gamma_{(i)a}}$ matrices, 
so to obtain an invariant
formulation of the Breit approximation for the electromagnetic
interaction. The final wave equation reads
\begin{eqnarray}
{}&{}\lambda=
q_a\Bigl( \breve{\gamma}_{(1)}\,{\gamma_{(1)}}_{a}-
\breve{\gamma}_{(2)}\,{\gamma_{(2)}}_{a}\Bigr)
+\Bigl(\,\breve{\gamma}_{(1)}
m_1+
\breve{\gamma}_{(2)}m_2\,\Bigr)+
{\displaystyle \frac \alpha r}
\spazio{1.2}\qquad\qquad\qquad\cr
{}&{}\qquad\qquad\qquad\qquad\qquad\qquad
+{\displaystyle \frac \alpha {2r}}\,
\breve{\gamma}_{(1)}\,\breve{\gamma}_{(2)}\,\Bigl( \,
{\gamma_{(1)}}_{a}\,{\gamma_{(2)}}_{a}
+
(n_a\,{\gamma_{(1)}}_{a})\,(n_a\,{\gamma_{(2)}}_{a})
\,\Bigr)
\label{Breit}
\end{eqnarray}
The difficulties of a direct solution of equation (\ref{Breit}) have 
been known for a long time (see \cite{Breit},\cite{FFT}): 
a perturbative interpretation of the term 
involving the vector potential, averaged over the pure Coulomb wave 
function -- very often at the non relativistic Schr\"odinger level of 
approximation --, is therefore what it is usually computed. 
The calculation can be found in \cite{Chi} 
and produces a perfect cancellation of the 
$\alpha^4$-perturbative Coulomb splitting. From the previous
discussion it is clear that the result should be tested numerically 
using the relativistic Coulomb 
wave functions for any value of the mass ratio. This numerical test
is however beyond the purpose of the present paper.
In any case -- and as it was certainly to be expected -- it appears 
that the
acceleration and delay effects of the Li\'enard-Wiechert potentials
-- and therefore the radiative corrections -- are  
essential for obtaining physical results.

\bigskip\bigskip
\vfill\break


\begin{Large}
\centerline{\bf Appendixes}
\end{Large}


   
\appendix\section{ The state vectors}


\renewcommand{\theequation}{A.\arabic{equation}}
\setcounter{equation}{0}

We give here the explicit form of the 16 component state vectors
of definite energy, angular momentum $(j,m)$ and even and odd
parity with respect to the angular momentum, namely $(-)^j$ and 
$(-)^{j+1}$. We call them $\Psi_{+}$ and $\Psi_{-}$ 
respectively. The even state  is then given by 
\begin{eqnarray}
\Psi_{+}={}^t\!\Bigl(\,
\Psi_{+}^{(M)},\,
\Psi_{+}^{(-M)},\,
\Psi_{+}^{(-\mu)},\,
\Psi_{+}^{(\mu)}
\,\Bigr)\,.
\label{StatoPari}
\end{eqnarray}
The four components is actually  multiplets composed by a singlet
and a triplet and will be indicated as:
\begin{eqnarray}
\Psi_{+}^{(\Lambda)}={}^t\!\Bigl(\,
\psi_{+\,0}^{(\Lambda)}\,,\,
\psi_{+\,1_+}^{(\Lambda)}\,,\,
\psi_{+\,1_0}^{(\Lambda)}\,,\,
\psi_{+\,1_-}^{(\Lambda)}
\,\Bigr)\,,
\end{eqnarray}
where $\Lambda=\pm M\,,\,\mp\mu\,$.
According to a common practice and in order to deduce real differential
equations in the following Appendix B,
we shall introduce an imaginary unity in front of the
coefficient functions $c_i(r)$ and $d_i(r)$
The explicit expressions of the state vector components 
are the following:
\begin{eqnarray} 
%
%
%
{}& \psi_{+\,0}^{(M)}= Y^{j}_{m}(\theta,\,\phi)\,\,{a_{0}}(r)
~~~~~~~~~\, 
\spazio {1.6}  \cr
%
%
{}& \psi_{+\,1_+}^{(M)}= - 
{\displaystyle \frac {\sqrt{j - m + 1}\sqrt{j + m}}{\sqrt{2\,j}
\sqrt{j+1}}}
Y^{j}_{m-1}(\theta,\,\phi)\,\,{b_{0}}(r)   
\spazio {1.6}  \cr
%
%
{}&\psi_{+\,1_0}^{(M)}= {\displaystyle \frac {m}{\sqrt{j}\,\sqrt{1 + j}}}
\,Y^{j}_{m}(\theta,\,\phi)\,\,
{b_{0}}(r) 
\spazio {1.6}  \cr
%
%
{}&\psi_{+\,1_-}^{(M)}= 
{\displaystyle \frac {
\sqrt{j - m}\,\sqrt{j + m + 1}}{\sqrt{2\,j}
\sqrt{j+1}}}
Y^{j}_{m+1}(\theta,\,\phi)\,\,{b_{0}}(r) 
\spazio {1.6}  \nonumber\cr
%
%
%
%
{}& \psi_{+\,0}^{(-M)}=Y^{j}_{m}(\theta,\,\phi)\,\,{a_{1}}(r)
\spazio {1.6}  \cr
%
%
{}& \psi_{+\,1_+}^{(-M)}=- {\displaystyle 
\frac {\sqrt{j - m + 1}\sqrt{j + m}}{\sqrt{2\,j}\sqrt{
 j+1}}}
Y^{j}_{m-1}(\theta,\phi)\,\,{b_{1}}(r)    
\spazio {1.6}  \cr
%
%
{}&\psi_{+\,1_0}^{(-M)}={\displaystyle \frac {m}{\sqrt{j}\,\sqrt{1 + j}}}
\,Y^{j}_{m}(\theta,\,\phi)\,
\,{b_{1}}(r)
\spazio {1.6}  \cr
%
%
{}&\psi_{+\,1_-}^{(-M)}= {\displaystyle \frac {\sqrt{j - m}\sqrt{j + m + 1}}{\sqrt{2\,j}
\sqrt{j+1}}}
Y^{j}_{m+1}(\theta,\phi)\,\,{b_{1}}(r)   
\spazio {1.6}  \nonumber\cr
%
%
%
%
%
%
{}& \psi_{+\,0}^{(-\mu)}=0
\spazio {1.6}  \cr
%
%
 {}& \psi_{+\,1_+}^{(-\mu)}={\displaystyle 
\frac {\sqrt{j + m
 - 1}\,\sqrt{j + m}}{\sqrt{2\,j}\,\sqrt{2\,j - 1}}}\,
 Y^{j-1}_{m-1}(\theta,\,\phi)\,
 i{c_0}(r)
+{\displaystyle \frac {\sqrt{j - m + 1}\,
 \sqrt{j - m + 2
}}{
\sqrt{ 2\,j+2}\,\sqrt{ 2\,j+3}}}
\,Y^{j+1}_{m-1}(\theta,\,\phi)\,i{d_{0}}(r) 
\spazio {1.6}  \cr
%
%
{}&\psi_{+\,1_0}^{(-\mu)}={\displaystyle \frac {\sqrt{j - m}\,\sqrt{j + m}}{\sqrt{j}\,
\sqrt{2\,j
 - 1}}}\,
Y^{j-1}_{m}(\theta,\,\phi)\,
i{c_0}(r)
- {\displaystyle \frac {\sqrt{j - m + 1}\,
 \sqrt{j + m + 1
}}{\sqrt{
1 + j}\,\sqrt{ 2\,j+3}}}\,Y^{j+1}_{m}(\theta,\,\phi)\,i{d_{0}}(r)  
\spazio {1.6}  \cr
%
%
{}&\psi_{+\,1_-}^{(-\mu)}={\displaystyle \frac {\sqrt{j - m
 - 1}\,\sqrt{j - m}}{\sqrt{2\,j}\,\sqrt{2\,j - 1}}}\,
 Y^{j-1}_{m+1}(\theta,\,\phi)\,
 i{c_0}(r)
+ {\displaystyle \frac {\sqrt{j + m + 1}\,
 \sqrt{j + m + 2
}}{
\sqrt{ 2\,j+2}\,\sqrt{ 2\,j+3}}}
\,Y^{j+1}_{m+1}(\theta,\,\phi)\,i{d_{0}}(r)   
\spazio {1.6}  \nonumber\cr
%
%
%
%
{}& \psi_{+\,0}^{(\mu)}=0  
\spazio {1.6}  \cr
%
%
{}&\psi_{+\,1_+}^{(\mu)}={\displaystyle 
\frac {\sqrt{j + m
 - 1}\,\sqrt{j + m}}{\sqrt{2\,j}\,\sqrt{2\,j - 1}}}\,
 Y^{j-1}_{m-1}(\theta,\,\phi)\,
 i{c_{1}}(r)
+ {\displaystyle \frac {\sqrt{j - m + 1}\,
\sqrt{j- m + 2
}}{
\sqrt{ 2\,j+2}\,\sqrt{ 2\,j+3}}}
\,Y^{j+1}_{m-1}(\theta,\,\phi)\,i{d_{1}}(r)   
\spazio {1.6}  \cr
%
%
 {}&\psi_{+\,1_0}^{(\mu)}={\displaystyle \frac {\sqrt{j - m}\,\sqrt{j + m}}{\sqrt{j}
 \,\sqrt{2\,j - 1
}}}\,
 Y^{j-1}_{m}(\theta,\,\phi)\,
 i{c_{1}}(r)
- {\displaystyle \frac {\sqrt{j - m + 1}\,
 \sqrt{j + m + 1
}}{\sqrt{
j + 1}\,\sqrt{ 2\,j+3}}}
\,Y^{j+1}_{m}(\theta,\,\phi)\,i{d_{1}}(r)   
\spazio {1.6}  \cr
%
%
{}&\psi_{+\,1_-}^{(\mu)}={\displaystyle 
\frac {\sqrt{j - m
 - 1}\,\sqrt{j - m}}{\sqrt{2\,j}\,\sqrt{2\,j - 1}}}\,
 Y^{j-1}_{m+1}(\theta,\,\phi)\,
 i{c_{1}}(r)
+ {\displaystyle \frac {\sqrt{j + m + 1}\,
\sqrt{j + m + 2
}}{
\sqrt{ 2\,j+2}\,\sqrt{ 2\,j+3}}}
\,Y^{j+1}_{m+1}(\theta,\,\phi)\,i{d_{1}}(r)
\nonumber\cr
%
%
%
%
\label{psipari16}
\end{eqnarray}
where $~Y^k_q(\theta,\,\phi)~$ are the spherical
harmonic functions. 
Concerning the state with opposite parity, it can be seen that the
action of the parity transformation is simply the exchange of
the first eight components with the second eight ones. Therefore,
in order to determine the differential equations describing the odd
states of the system, we shall use the
state $\Psi_{-}$ given by
\begin{eqnarray}
\Psi_{-}=
%
%
%
%
\left(\,
\begin{matrix}
0 & {\bf I}_8  \spazio{0.8}\cr
 {\bf I}_8 & 0 
 \end{matrix}
 \,\right)\,\Psi_{+}
%
%
%
\label{StatoDispari}
\end{eqnarray}
As observed in Section IV.B this amounts to changing the sign of the
mass $m_{1}$.
%
%

\section{ The original system} 
%
%
%
\renewcommand{\theequation}{B.\arabic{equation}}
\setcounter{equation}{0} 

As we said in Section (IV.A), in order to determine the system of
differential equations describing the two body problem, we
shall apply the Hamiltonian operator $H_0$,
given in (\ref{H}), to the state vector  (\ref{StatoPari}). In each
component we require the vanishing of the coefficient of each different
spherical harmonics. This leads to a system of thirty-four
differential equations  in
the unknown functions  $\,a_i(r)\,,\, b_i(r)\,,\, c_i(r)\, 
{\mathrm {and}}~~
d_i(r)\,,\, (i=0,1)\,,$ that appear in the state vector. However, 
as expected, only eight of these differential equations are
independent. They read
\begin{subequations}
\label{original_system} 
\begin{eqnarray}
&{}&\sqrt{j}\,\Bigl({\displaystyle\frac {d}{dr}}
 + {\displaystyle \frac {j + 1}{r}\Bigr)\,({a_{0}}(r) + {a_{1}}(r)
)} - \sqrt{j + 1}\,\Bigl({\displaystyle\frac {d}{dr}} + 
{\displaystyle \frac {j + 1}{r}\Bigr)\,({b
_{0}}(r) - {b_{1}}(r))}
\spazio{1.2}\cr
&{}&\qquad\qquad\qquad\qquad\qquad\qquad\qquad\qquad\qquad\qquad
 +\,\sqrt{2\,j + 1}\,{c_{0}}(r)\,(\mu  + h(r))=0~~~~~ 
 \spazio{1.2}\\
&{}&\sqrt{j}\,\Bigl({\displaystyle\frac {d}{dr}}
 + {\displaystyle \frac {j + 1}{r}\Bigr)\,({a_{0}}(r) + {a_{1}}(r)
)} + \sqrt{j + 1}\,\Bigl({\displaystyle\frac {d}{dr}} + 
{\displaystyle \frac {j + 1}{r}\Bigr)\,({b
_{0}}(r) - {b_{1}}(r))}
\spazio{1.2}\cr
&{}&\qquad\qquad\qquad\qquad\qquad\qquad\qquad\qquad\qquad\qquad
 -\,\sqrt{2\,j + 1}\,{c_{1}}(r)\,(\mu  - h(r))=0~~~~~
 \spazio{1.2}\\
&{}&\sqrt{j + 1}\,\Bigl({\displaystyle\frac {d}{dr}}
 - {\displaystyle \frac {j}{r}\Bigl)\,({a_{0}}(r) + {a_{1}}(r))
} + \sqrt{j}\,
\Bigl({\displaystyle\frac {d}{dr}}
 - {\displaystyle \frac {j}{r}\Bigl)\,({b_{0}}(r) - {b_{1}
}(r))}
\spazio{1.2}\cr
&{}&\qquad\qquad\qquad\qquad\qquad\qquad\qquad\qquad\qquad\qquad
-\,\sqrt{2\,j + 1}\,{d_{0}}(r)\,(\mu  + h(r)
)=0~~~~~ 
\spazio{1.2}\\
&{}&\sqrt{j + 1}\,\Bigl({\displaystyle\frac {d}{dr}}
 - {\displaystyle \frac {j}{r}\Bigl)\,({a_{0}}(r) + {a_{1}}(r))
}  - \sqrt{j}\,
\Bigl({\displaystyle\frac {d}{dr}}
 - {\displaystyle \frac {j}{r}\Bigr)\,({b_{0}}(r) - {b_{1}
}(r))}
\spazio{1.2}\cr
&{}&\qquad\qquad\qquad\qquad\qquad\qquad\qquad\qquad\qquad\qquad
+\, \sqrt{2\,j + 1}\,{d_{1}}(r)\,(\mu  - h(r)
)=0~~~~~
\spazio{1.2}\\
&{}&\sqrt{j}\,\Bigl({\displaystyle\frac {d}{dr}}
 - {\displaystyle \frac {j - 1}{r}\Bigr )\,({c_{0}}(r) + {c_{1}}(r)
)}
- \sqrt{j + 1}\,\Bigl({\displaystyle\frac {d}{dr}}
 + {\displaystyle \frac {j + 2}{r}\Bigr)\,({d
_{0}}(r) + {d_{1}}(r))} 
\spazio{1.2}\cr
&{}&\qquad\qquad\qquad\qquad\qquad\qquad\qquad\qquad\qquad\qquad
 + \,\sqrt{2\,j + 1}\,{a_{0}}(r)\,(M - h(r))=0~~~~
 \spazio{1.2}\\
&{}&\sqrt{j}\,\Bigl({\displaystyle\frac {d}{dr}}
 - {\displaystyle \frac {j - 1}{r}\Bigr )\,({c_{0}}(r) + {c_{1}}(r)
)}
- \sqrt{j + 1}\,\Bigl({\displaystyle\frac {d}{dr}} + 
{\displaystyle \frac {j + 2}{r}\Bigr )\,({d
_{0}}(r) + {d_{1}}(r))}
\spazio{1.2}\cr
&{}&\qquad\qquad\qquad\qquad\qquad\qquad\qquad\qquad\qquad\qquad
 - \,\sqrt{2\,j + 1}\,{a_{1}}(r)\,(M + h(r))=0~~~~~
 \spazio{1.2}\\
&{}&\sqrt{j + 1}\,\Bigl({\displaystyle\frac {d}{dr}}
 - {\displaystyle \frac {j - 1}{r}\Bigr)\,({c_{0}}(r) - 
{c_1}(r))} 
+ \sqrt{j}\,\Bigl({\displaystyle\frac {d}{dr}}
 + {\displaystyle \frac {j + 2}{r}\Bigr)\,({d_{0}}(r) - 
{d_1}(r))}
\spazio{1.2}\cr
&{}&\qquad\qquad\qquad\qquad\qquad\qquad\qquad\qquad\qquad\qquad
- \,\sqrt{2\,j + 1}\,{b_{0}}(r)\,(M - 
h(r)) =0 ~~~~~
\spazio{1.2}\\
&{}&\sqrt{j + 1}\,\Bigl({\displaystyle\frac {d}{dr}}
 - {\displaystyle \frac {j - 1}{r}\Bigl)\,({c_{0}}(r) - 
{c_1}(r))} 
+ \sqrt{j}\,\Bigl({\displaystyle\frac {d}{dr}} 
+ {\displaystyle \frac {j + 2}{r}\Bigl )\,({d_{0}}(r) - 
{d_1}(r))}
\spazio{1.2}\cr
&{}&\qquad\qquad\qquad\qquad\qquad\qquad\qquad\qquad\qquad\qquad
- \,\sqrt{2\,j + 1}\,{b_{1}}(r)\,(M + 
h(r)) =0 ~~~~~
\end{eqnarray}
\end{subequations}

\smallskip

Once the definitions (\ref{variabilipm}) and (\ref{varuvpm}) are used,
a straightforward, although somewhat lengthy computation, leads directly
to (\ref{relalgeven}) and (\ref{diffeqeven}).


\section{ A perturbative calculation in the mass ratio }


\renewcommand{\theequation}{C.\arabic{equation}}
\setcounter{equation}{0}

We now present a perturbative approach to the ground level 
of the Hydrogen atom, taking as perturbation parameter the quantity
\begin{eqnarray}
\epsilon=1-\rho_H=1-{\displaystyle\frac {m_P-m_e}{m_P+m_e}}=0.0010886411
\end{eqnarray}
As we have said in Section \ref{Sec_spectrum}, for $j=0$ the system splits in
two subsystems obtained by (\ref{sistema_mat}-\ref{sysmat}), or,
equivalently, by the second order equation (\ref{U0pari_1}). 
For later convenience in the development of the perturbative 
calculations we introduce
the modified unknown functions $\phi(r),\,\theta(r)$ defined by
\begin{eqnarray}
y_1(r)=\,{\displaystyle\frac {\sqrt{F(r)}}r} \,\phi(r)\,,
\qquad y_2(r)=\,{\displaystyle\frac {\sqrt{F(r)}}r}\,\theta(r)\,.
\label{phitheta}
\end{eqnarray}
In particular we see that second order equation (\ref{U0pari_1}) written 
in terms of $\phi(r)$ does not present the first derivative term:
\begin{eqnarray}
&{}& {\displaystyle\frac {d^{2}}{dr^{2}}}\,\phi(r) +  \Bigl(  \, 
{\displaystyle \frac {1}{2}} \,{\displaystyle \frac 
{{\displaystyle\frac {d}{d
r}}\,(r^{2}\,{\displaystyle\frac {d}{dr}}\,{F}(r))}{r^{2}\,{F}
(r)}}  - {G}(r)\,{F}(r) - 
{\displaystyle \frac {3}{
4}} \,{\displaystyle \frac {({\displaystyle\frac {d}{dr}}\,{F}(r))^{2}
}{{F}^{2}(r)}}  \,  \Bigr) \,\phi(r) = 0
\spazio{1.2}
\label{phi2senza}
\end{eqnarray}
For $F(r)$ and $G(r)$ we shall then assume their Coulomb explicit form 
(\ref{equaUVW}). Finally,
the parameters are rescaled according to 
(\ref{massrescale}-\ref{h_x}) and, correspondingly, 
we write the series expansions of the eigenvalue and the eigenfunction
components in terms of  $\epsilon\,$:
\begin{eqnarray}
&{}&\lambda=\sum_n\,\epsilon^n\,\lambda_n\,,\qquad\quad
y_j(x)=\sum_n\,\epsilon^n\,y_j^{(n)}(x)\,,~~~(j=1\,,3
\,)\,.~~
\end{eqnarray}

At the lowest order in $\epsilon$ we have the system
\begin{eqnarray}
&{}& {\displaystyle\frac {d}{dx}}\,{y_{1}^{(0)}}(x) - (1 + {\lambda _{0}} + 
{\displaystyle \frac {\alpha }{x}} )\,{y_{3}^{(0)}}(x)=0
\spazio{1.2}\cr
&{}& {\displaystyle\frac {d}{dx}}\,{y_{3}^{(0)}}(x) +
 {\displaystyle \frac {2}{x}}\,{y_{3}^{(0)}}(
x)  + ({\lambda _{0}} - 1 + {\displaystyle \frac {\alpha }{x
}} )\,{y_{1}^{(0)}}(x)=0\cr
&{}& {}
\end{eqnarray}
in which we obviously recognize the Dirac equation for $j=0$.
The solution of the system have therefore the well known expressions,
\cite{Landau},
\begin{eqnarray}
&{}& {y_{1}^{(0)}}(x)={N_0}\,\sqrt{1 + {\lambda _{0}}}\,
\,e^{ - \frac {\xi }{2}}
\,\xi ^{\sqrt{1 - \alpha ^{2}} - 1}\,(\,{}_1\!F_1(a,b,\xi) + A\,
{}_1\!F_1(a+1,b,\xi)\,)
\spazio{1.2}\cr
&{}& {y_{3}^{(0)}}(x)= - {N_0}\,\sqrt{ - {\lambda _{0}} + 1}
\,\,e^{ - \frac {\xi 
}{2}}
\xi ^{\sqrt{1 - \alpha ^{2}} - 1}\,
(\,{}_1\!F_1(a,b,\xi) - A\,
{}_1\!F_1(a+1,b,\xi)\,)
\label{solu_y}
\end{eqnarray}
where 
\begin{eqnarray}
&{}& \xi =2\,\sqrt{1 - {\lambda _{0}}^{2}}\,x\,,\qquad
A={\displaystyle \frac {\sqrt{1 - \alpha ^{2}}\,\sqrt{1 - {
\lambda _{0}}^{2}} - {\lambda _{0}}\,\alpha }{\alpha  + \sqrt{1
 - {\lambda _{0}}^{2}}}}\cr
 &{}& {}
 \end{eqnarray}
while the parameters of the confluent hypergeometric functions are
\begin{eqnarray}
&{}& {a}=\sqrt{1 - \alpha ^{2}} - {\displaystyle \frac {{\lambda 
_{0}}\,\alpha }{\sqrt{1 - {\lambda _{0}}^{2}}}}\,,\qquad
{b}=1 + 2\,\sqrt{1 - \alpha ^{2}}\,.~~~~~~~
\end{eqnarray}
Finally ${N_0}$ is the normalizing constant that 
obviously makes sense only
when assuming the spectral condition
\begin{eqnarray}
\lambda_0=
\Bigl(\,1 + {\displaystyle \frac {\alpha ^{2}}{
(\,n + \sqrt{1 - \alpha ^{2}
}\,)^{2}}}\,\Bigr)^{-\frac 12}
\end{eqnarray}
with integer $n$. Letting
\begin{eqnarray}
\lambda_0=1+\alpha^2\,w_0\,,
\end{eqnarray}
the value for the ground state of the Dirac equation is
$w_0=-.5000066566$.
 
We now switch the perturbative machine on. In the first place we
observe that, according to (\ref{phitheta}), 
the functions $\phi^{(0)}(x)$ and $\theta^{(0)}(x)$ are just obtained 
by the product with 
the limiting value of the rescaled multiplicative factor, {\it i.e.}
$x^{-1}\,(1+\lambda _{0} +\alpha/x)^{1/2}$. In particular, 
$\phi^{(0)}(x)$ solves the second order equation
\begin{eqnarray}
&{}&{\displaystyle\frac {d^{2}}{dx^{2}}}\,{\phi ^{(0)}}(x) -  \Bigl(  \, 
{\displaystyle \frac {x^{2} - (x\,{\lambda _{0}} + \alpha )^{2}}{
x^{2}}}  +
{\displaystyle \frac {3}{4}} \,{\displaystyle \frac {
\alpha ^{2}}{(x + x\,{\lambda _{0}} + \alpha )^{2}\,x^{2}}}   
 \Bigr) \,{\phi ^{(0)}}(x)=0~~~~~~~~
\label{e2phi0}
\end{eqnarray}
We then go to the linear terms in the expansion parameter $\epsilon$ 
and, after some simple
calculations we find a second order equation in  ${\phi _{1}}(x)$
involving $\lambda_1$ and the zeroth order quantities. As it should be, 
according to the standard scheme of perturbation theory, 
the second order self-adjoint operator acting on 
${\phi _{1}}(x)$ is the same as the one acting on ${\phi _{0}}(x)$ 
in the zeroth order equation (\ref{e2phi0}). Indeed:
\begin{eqnarray}
&{}&{\displaystyle\frac {d^{2}}{dx^{2}}}\,{\phi ^{(1)}}(x) - 
\Bigl( {\displaystyle \frac {  x^{2} - ( x\,{
\lambda _{0}}+\alpha )^{2}}{x^{2}}} 
+ {\displaystyle 
\frac {3}{4}} \,{\displaystyle \frac {\alpha ^{2}}
{(x + x\,{\lambda _{0}} + \alpha)^{
2}\,x^{2}}} \Bigl)\,{\phi ^{(1)}}(x)\, - 
\spazio{1.4}\cr
&{}& 
\Biggl[\,{\displaystyle \frac 1{x^3}}\,
\Bigl(\,{\displaystyle \frac 
{({3\,x}  + x\,{\lambda _{0}} + 
\alpha )^{2}}{4}}\,
{- 2\,x\,(x + x\,{\lambda _{0}} + \alpha) - x
^{2}\,{\lambda _{1}}\Bigr)\,(x + x\,{\lambda _{0}} + \alpha)\,
{\phi ^{(0)}}(x)}  + 
\spazio{1.4}\cr
&{}& \qquad\qquad\qquad\qquad\qquad\qquad
\Bigl(
 {\lambda _{1}} -  \,{\displaystyle \frac 
 {(x + x\,{\lambda _{0}} + \alpha)^{2}}{4\,x^{2}}}  
 \Bigr)\,{\displaystyle\frac {d}{dx}}\,{\theta 
^{(0)}}(x)+ 
\spazio{1.4}\cr
&{}& \Bigl(\,\lambda_1\,\Bigl(\, \,{\displaystyle \frac {
\alpha }{2x\,(x + x\,{\lambda _{0}} + \alpha)}}  + 
{\displaystyle \frac {1}{x}}\, \Bigr)\, -
 {\displaystyle \frac {1}{8\,x^{3}}} \, 
{(x + x\,{\lambda _{0}} + \alpha)\,( - \alpha  + 2\,x + 2\,x\,
{\lambda _{0}})}
 \Bigr)\,{\theta ^{(0)}}(x)\,\Biggr]=0\cr
&{}& 
\label{e2phi1}
\end{eqnarray}
Therefore the correction $\lambda_1$ can
be calculated from equation  (\ref{e2phi1}) in the standard way: 
we take for ${\phi _{0}}(x)$ and ${\theta _{0}}(x)$ the
solution obtained by (\ref{solu_y}) and the spectral condition, we
multiply (\ref{e2phi1}) by ${\phi _{0}}(x)$
and integrate the result from zero to infinity. Letting then
$\lambda_1=\alpha^2\,w_1$ and carrying over the numerical calculations
we have indicated, we obtain the value
$w_1= .232995\, 10^{-4}$, so that the correction to the ground
state of the Hydrogen results in 
$(1-\rho_H)\,w_1= 0.253509\, 10^{-7}$. This result is in excellent
agreement with what is found from the numerical integration, as 
presented in Section \ref{Sec_results}. When applied to Muonium, 
\cite{Jun}, where we have $(1-\rho_{\mathrm{muonium}})=0.0096261$, the
correction would amount to $0.22428\, 10^{-6}$.


\section{ The numerical scheme}


\renewcommand{\theequation}{D.\arabic{equation}}
\setcounter{equation}{0}

We outline the numerical methods used to
integrate the even and odd fourth-order differential systems 
and to find the spectrum. For the numerical treatment we have rewritten the mass parameters 
in the systems according to the definitions 
(\ref{massrescale}-\ref{h_x}) and 
we have adopted the independent variable $x=m_e\,r$.

As a first remark, we see that both systems 
are singular at zero and infinity. Therefore,
in order to start the numerical integration, the appropriate 
asymptotic solutions
at the two singular points must be determined by the appropriate 
developments. We report here below and separately on some further 
details  
of these solutions. As a general feature, 
however, it turns out that among
the four independent solutions that can be determined at each singular 
point, two of them are divergent and would produce not square-integrable
wave functions, {\it i.e.} outside of the Hilbert 
space where we want to solve the spectral problem. 
These solutions will therefore be rejected. 

Once the two acceptable solutions in zero and infinity have 
been determined,
the eigenvalue will be obtained by adapting to our problem the well
known ``double shooting method". 
More precisely, for each of the acceptable solutions both
at zero and at infinity we start the numerical integration up to a 
chosen crossing point $x_c$. In $x_c$ we impose the matching conditions
for each component of the wave function. These give rise to the
following linear system:
\begin{eqnarray}
&{}& 
K_{1}\,{y^{(0,1)}_j}({x_c}) + K_{2}\, y^{(0,2)}_j({x_c})
= 
K_{3}\,y^{(\infty,1)}_j({x_c}) + 
K_{4}\,y^{(\infty,2)}_j({x_c})
\label{sistemaCK}
\end{eqnarray}
where $j=1,4$.
The first superscripts ``$\,0\,$" and ``$\,\infty\,$" 
evidently refer to
the singular point where the integration started, while the 
second subscripts
``$\,1\,$" and ``$\,2\,$" denote the solution chosen at each 
boundary. 
Being linear
and homogeneous, the  system (\ref{sistemaCK}) has nontrivial
solutions in $K_1,\,K_2,\,K_3$ and $K_4$ provided its determinant
is vanishing. The condition

\begin{eqnarray}
{\det} \left( \! 
{\begin{array}{cccc}
{y^{(0,1)}_1}({x_c}) & y^{(0,2)}_1({x_c}) &
y^{(\infty,1)}_1({x_c}) & y^{(\infty,2)}_1({x_c}) 
\spazio {1.4}  \\
{y^{(0,1)}_2}({x_c}) & y^{(0,2)}_2({x_c}) &
y^{(\infty,1)}_2({x_c}) & y^{(\infty,2)}_2({x_c}) 
\spazio {1.4}  \\
{y^{(0,1)}_3}({x_c}) & y^{(0,2)}_3({x_c}) &
y^{(\infty,1)}_3({x_c}) & y^{(\infty,2)}_3({x_c}) 
\spazio {1.4}  \\
{y^{(0,1)}_4}({x_c}) & y^{(0,2)}_4({x_c}) &
y^{(\infty,1)}_4({x_c}) & y^{(\infty,2)}_4({x_c}) 
\end{array}}
  \! \right)\! =0\cr
\label{determinante}  
\end{eqnarray}
\smallskip

\noindent
gives therefore the equation that has to be solved in order to find the
spectral values $\lambda$. Needless to say that when the system
decouples in two separate second order differential equations, no
additional modification to the usual double shooting method is necessary
and each equation has been solved independently.

We now describe  the behavior at the singular points of the two 
systems.

\subsection{The behavior at zero}

We shall first consider the behavior of the systems at the origin by
expanding the equations of each system in a neighborhood of $x=0$.
Accordingly, the solutions will be assumed of the form
\begin{eqnarray}
{y_{i}}(x)\,=\,x^{\nu }~{\displaystyle \sum _{j=0}^{N}} ~~{y_{i}^{(j)}}
\,x^{j}\,,\qquad(i=1,4)
\label{serie_y}
\end{eqnarray}
The procedure is the standard one. We substitute (\ref{serie_y}) 
into the system (\ref{sistema_mat}) and take the different orders in
$x$. The lowest order gives the indicial equations that produce four
different values for the exponent $\nu$, only two of which
are acceptable. In correspondence to the indices, some relations 
for the ${y_{i}^{(0)}}$ are established. It also turns out that the
indices are equal both for the even and the odd systems. 
We have a first index
\begin{eqnarray}
\nu_1= - 1 + \sqrt{1 -  {\displaystyle\frac  {\alpha ^{2}}4}  + 
j\,(1 + j) }
\label{nu_1}
\end{eqnarray}
with corresponding relations
\begin{eqnarray}
{y_{1}^{(0)}}=1\,,\qquad
{y_{3}^{(0)}}= 
{\displaystyle \frac {2}{\alpha}}\,{\nu_1 } \,,\qquad
{y_{2}^{(0)}}={y_{4}^{(0)}}=0\,~
\label{rel_nu_1}
\end{eqnarray}
and a second index (only for $j>0$)
\begin{eqnarray}
\nu_2= - 1 + \sqrt{j\,(1 + j)  
- {\displaystyle \frac {\alpha ^{2}}{4}} }
\label{nu_2}
\end{eqnarray}
for which
\begin{eqnarray}
{y_{1}^{(0)}}={y_{3}^{(0)}}=0\,,\qquad
{y_{2}^{(0)}}=1\,,\qquad
{y_{4}^{(0)}}=-{\displaystyle \frac {2}{\alpha}}\,(1+{\nu_2 })
\label{rel_nu_2}
\end{eqnarray}

The corresponding (very cumbersome)
series have been calculated formally up to the eight order
and have been tested by substitution in the differential equations
with the help of a symbolic computer package. They have also been
tested numerically, with physical values of the parameters and 
varying value of the coordinate $x$. In addition, 
this procedure has furnished an effective criterion
for choosing the boundary point near zero: in fact we have assumed as
acceptable a value of $x$ where the ratio of the differential
equation evaluated on the solution over the solution itself 
was at most of the order of $10^{-12}$.

\subsection{The behavior at infinity}

In order to determine the solutions in the neighborhood of infinity
we consider the following asymptotic expansion for the
components of the solution must be looked in the form (see \cite{Cope}) 
\begin{eqnarray}
{y_{i}}(x)=e^{ - \gamma \,x}\,x^{\nu }~ {\displaystyle \sum 
_{j=0}^{N}} ~~{y_{i}^{ (j)}}\,x^{ - j} ,~~(\,i=1,4\,)
\end{eqnarray}
Contrary to what occurs in zero, also the indices of the 
asymptotic expansion
depend upon the eigenvalue $\lambda$. At the lowest order, indeed,
we find 
\begin{eqnarray}
\gamma ^{2}={\displaystyle \frac {1 + \rho ^{2}}{(1 - \rho )^{2}}
}  - {\displaystyle \frac {\lambda ^{2}}{4}}  - {\displaystyle 
\frac {4\,\rho ^{2}}{\lambda ^{2}\,(1 - \rho )^{4}}}
\end{eqnarray}
and we shall take the positive square root to insure the convergence.
The next order determines the exponent $\nu$, given by
\begin{eqnarray}
\nu=-1+{\displaystyle\frac 1{2\,\gamma}}\,\Bigl(\,   
 {\displaystyle \frac {\alpha \,
\lambda}{2}}  - {\displaystyle \frac {8\,\rho ^{2}\,\alpha }{h^{3}\,(
  1 - \rho )^{4}}}
\,\Bigr)
\end{eqnarray}
The different choices of the two coefficients $y_1^{(0)}$ and 
$y_2^{(0)}$ determine then two different series and therefore two 
linearly independent asymptotic solutions of the system.

\subsection{The wave function}
\label{WF}

Once the spectrum has been found, we can proceed to determine the
wave function. If $\lambda_\ast$ is one of the spectral values, then,
in the system (\ref{sistemaCK}) we can fix one of the four parameters
$K_i$ to unity and solve for the remaining ones. For instance for
the Hydrogen atom and the $2p_{1/2}$ state of the Dirac spectroscopy, 
for which
$\lambda_\ast=-0.1250020774$, we find
\begin{eqnarray}
K_1=40.26667371\,,\qquad K_2=.6304491167\,,\qquad
K_3=.7071066401\,,\qquad K_4=1\,.
\end{eqnarray}
Therefore the complete wave function will result in 
\begin{eqnarray}
&{}&y_j(x)=K_{1}\,{y^{(0,1)}_j}({x}) + K_{2}\, y^{(0,2)}_j({x})\,,\qquad
~~\,{\mathrm {for}}~~~x<x_c\cr
\spazio{1.6}
&{}&\phantom{y_j(r)}=
K_{3}\,y^{(\infty,1)}_j({x}) + 
K_{4}\,y^{(\infty,2)}_j({x})\,,\qquad
{\mathrm {for}}~~~x>x_c
\end{eqnarray}
The plot of $y_1(x)$ ig given in the following figure:

\medskip\medskip\medskip\medskip\medskip

\begin{center}
\includegraphics[height=6.cm,width=7cm]{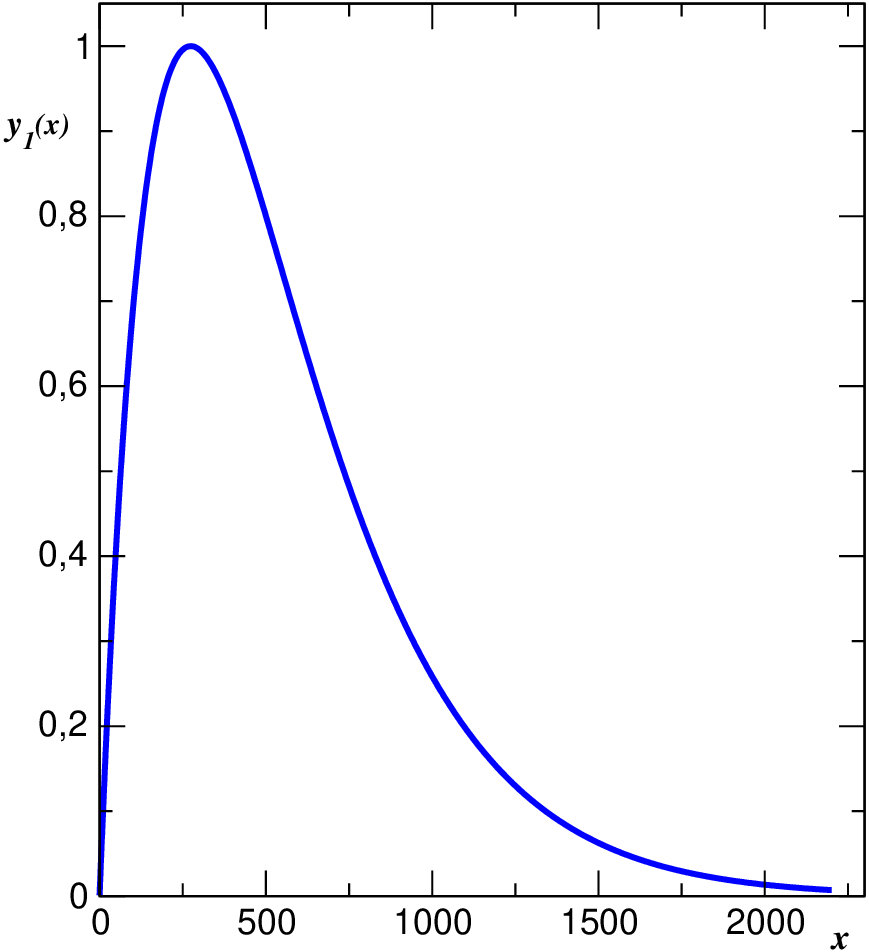}
\end{center}
\medskip

\begin{footnotesize}
\noindent{ FIG. 3 The first component  of the wave function of the
Coulomb Hydrogen atom for the $w_{\,+,\,\,j\,=\,1,\,\,{\mathrm I}}$ 
state.
}
\end{footnotesize}


\section{ The ground state  wave functions   of Parapositronium 
and  Orthopositronium}


\renewcommand{\theequation}{E.\arabic{equation}}
\setcounter{equation}{0}

In this last appendix we give some details on the two wave functions
of the ground state of the Parapositronium and  Orthopositronium 
corresponding to the states $w_{\,+,\,j\,=\,0,\,\,{\mathrm {I}}}$  and 
$w_{\,-,\,j\,=\,1,\,\,{\mathrm {I}}}\,$, degenerate in energy.
We finally comment on the calculation of the hyperfine splitting of the
Positronium ground level.

\subsection{The Parapositronium}

The case of the Parapositronium, namely the state with even parity, 
is comparatively easy,
since it can be solved by discussing the single second order
differential equation (\ref{U0pari_1}) for the Coulomb
potential, with $\rho=0$ and $j=0$. 
Its explicit form reads:
\begin{eqnarray}
{\frac {d^{2}}{dx^{2}}}\,{y_1}(x) + \Bigl({\displaystyle 
\frac {2}{x}}  + {\displaystyle \frac {2\,\alpha }{x\,((4 + w\,
\alpha ^{2})\,x + 2\,\alpha )}} \,\Bigr)\,{\frac {d}{dx}}\,{y_1}(x
) + \Bigl(\,{\displaystyle \frac {((4 + w\,\alpha ^{2})\,x + 2\,\alpha 
)^{2}}{16\,x^{2}}}  - 1\,\Bigr)\,{y_1}(x) 
\label{equaposiparifd1}
\end{eqnarray}

Zero and infinity are the two singular points of this equation. At
infinity there is a divergent and a convergent solution, so that
the latter will be assumed. In zero both solutions are divergent, 
and only one of them, the solution we shall accept, is 
normalizable. The regular
solution in zero has the following expansion: 
\begin{eqnarray}
{y_1}(x)\,
{\phantom{x}\atop{\simeq\atop{x\rightarrow 0}}}\,
A\,x^{\displaystyle{\Bigl( - 1 + \frac {\sqrt{4 - \alpha
^{2}}}{2}}\Bigr)}\,\Bigl(\,1
 - {\displaystyle \frac {(\alpha ^{2} + 2 - \sqrt{4 - \alpha ^{2}
})\,(4 + w\,\alpha ^{2})}{\alpha \,(4\,\sqrt{4 - \alpha ^{2}}
 + 4)}}\,x \,\Bigr)\,,
\end{eqnarray}
where $A$ is an arbitrary integration constant.

The asymptotic behavior of the
solution by taking the dominant term of (\ref{equaposiparifd1}), 
at infinity. The solution of the corresponding asymptotic equation 
reads
\begin{eqnarray}
{y_1}(x)\,
{\phantom{x}\atop{\simeq\atop{x\rightarrow \infty}}}\,B\,
\exp\Bigl(\,-{\displaystyle \frac \alpha 2}\,\sqrt{ - w}\,\sqrt{2+w\,
{\displaystyle\frac {\alpha^2}4}}\,x\,\Bigr)
\end{eqnarray}
where $B$ is an arbitrary integration constant. However for
calculating a better approximation to the asymptotic boundary conditions
we can take the expansion of (\ref{equaposiparifd1})
up to terms of order $x^{-2}$. In fact the resulting equation 
can be exactly solved in terms of Whittaker $M$ and $W$ functions. 
Up to the usual integration constant we can fix equal to unity, 
the solution with regular behavior at infinity is:
\begin{eqnarray}
{y_1}(x)
{\phantom{x}\atop{\simeq\atop{x\rightarrow \infty}}}
{\displaystyle\frac{\sqrt{(4 + w\,\alpha ^{2})\,x
 + 2\,\alpha }}{x^{3/2}}}\,\,
{W}\Bigl(\,{\displaystyle \frac {4 + w\,\alpha ^{2}}{2\,
\sqrt{ - w}\,\sqrt{w\,\alpha ^{2} + 8}}} , \,{\displaystyle 
\frac {\sqrt{1 - \alpha ^{2}}}{2}} , \,{\displaystyle \frac {
\alpha \,\sqrt{ - w}\,\sqrt{w\,\alpha ^{2} + 8}}{2}}\,x\, \Bigr)\,.
\end{eqnarray}

Once the boundary behaviors are established, the wave function 
component $y_1(x)$ and its first derivative $y'_1(x)$ 
are computed by standard numerical integration. 
Finally, in order
to determine the complete sixteen component wave function, 
we only have to
calculate the eight coefficient functions $a_i(x),b_i(x),c_i(x)$ and 
$d_i(x)$, $i=0,1$, in terms of $y_1(x)$ and its first derivative 
$y'_1(x)$. 
Recalling that $y_2(x)=0$, and that from (\ref{sistema_mat}) and
(\ref{sysmat})we have
\begin{eqnarray}
{y_{3}}(x)={\displaystyle \frac {2\, x\,
y'_1(x)}{\alpha  + 
x\,(4 + {w\,\alpha ^{2}} )}}
\end{eqnarray}
simple substitutions lead to the following expressions:
\begin{eqnarray}
{}& a_0(x)=\Bigl({\displaystyle \frac {1}{2}}  + {\displaystyle 
\frac {2\,x}{x\,(4 + w\,\alpha ^{2}) + 2\,\alpha }}  \Bigr)\,{y_{1}}(x)
\,,
\qquad
a_1(x)=\Bigl({\displaystyle \frac {1}{2}}  - 
 {\displaystyle \frac {2\,x}{x\,(
4 + w\,\alpha ^{2}) + 2\,\alpha }}  \Bigr)\,{y_{1}}(x)\,, 
\spazio{1.6}\cr
{}& b_0(x)=b_1(x)=c_0(x)=c_1(x)=0\,,\qquad
d_0(x)=d_1(x)={\displaystyle \frac 12}\,y_3(x)\,.
\end{eqnarray}
The components $y_1(x)$ and $y_3(x)$ are respectively represented 
in the first plot of Fig. 4.

\subsection{The Orthopositronium}

Let us then consider the wave function for the odd state. The system to
be discussed is now given by (\ref{sistema_mat} - \ref{sysmat} -
\ref{equaUVW_C}) with $\rho=0$ and $j=1$.
Considering the series expansion in the origin, we see
that we have two regular solutions with indices
\begin{eqnarray}
\nu_1=\sqrt{3-\frac{\alpha^2}{4}}-1\,,~~~~~~~~~~~~~~~~~~~
\nu_2=\sqrt{2-\frac{\alpha^2}{4}}-1\,.
\end{eqnarray}
On the other side, the asymptotic expansion gives then solutions 
presenting the exponential decrease
\begin{eqnarray}
\exp\Bigl(\,-{\displaystyle \frac \alpha 2}\,\sqrt{ - w}\,\sqrt{2+w\,
{\displaystyle\frac {\alpha^2}4}}\,\,x\,\Bigr)\,,
\end{eqnarray}
exactly the same as for Parapositronium.

By applying the matching procedures described in Appendix \ref{WF}
we find the four components $y_i(x)$. They given in the second
plot Fig.4. 

Letting $S$ be the matrix
\begin{eqnarray}
S={\displaystyle\frac{2\sqrt{6}}{(4+\alpha^2 w)x+2\alpha}}
\left( 
{\begin{array}{cccc}
 -\sqrt{2} & 1 & -{\displaystyle\frac{(8+\alpha^2 w)x+2\alpha}
 {4\sqrt{2}}} & -{\displaystyle\frac{(8+\alpha^2 w)x+2\alpha}4}
 \spazio{1.2}\cr  
 -\sqrt{2} & -1 &-{\displaystyle\frac{\alpha}
 {4\sqrt{2}}}\,(2+\alpha w)x & {\displaystyle\frac{\alpha}
 {4}}\,(2+\alpha w)x
 \spazio{1.2}\cr  
 -1 & -\sqrt{2} & {\displaystyle\frac{(8+\alpha^2 w)x+2\alpha}
 {4}} & {\displaystyle\frac{(8+\alpha^2 w)x+2\alpha}
 {4\sqrt{2}}} 
 \spazio{1.2}\cr  
 -1 & \sqrt{2} & {\displaystyle\frac{\alpha}
 {4}}\,(2+\alpha w)x & {\displaystyle\frac{\alpha}
 {4\sqrt{2}}}\,(2+\alpha w)x 
\end{array}}
 \right) 
\end{eqnarray}
the explicit relationships among the four components $y_k(x)$ and the 
eight coefficients of the complete state vector, 
$a_i(x),\,b_i(x),\,c_i(x),\,d_i(x),\,$ $i=0,1$  are the following:
\begin{eqnarray}
a_0(x)=a_1(x)={\displaystyle \frac {1}{2}} \,{y_1}(x)\,,\qquad
b_0(x)=-b_1(x)={\displaystyle \frac {1}{2}} \,{y_2}(x)
\end{eqnarray}
and
\begin{eqnarray}
{}^t\!(c_0(x),c_1(x),d_0(x),d_1(x))\,=\,
S\,\,\,{}^t\!(y_1(x),y_2(x),y_3(x),y_4(x))\,.
\end{eqnarray}

\medskip\medskip\medskip\medskip\medskip
\begin{center}
\includegraphics[height=6.cm,width=6cm]{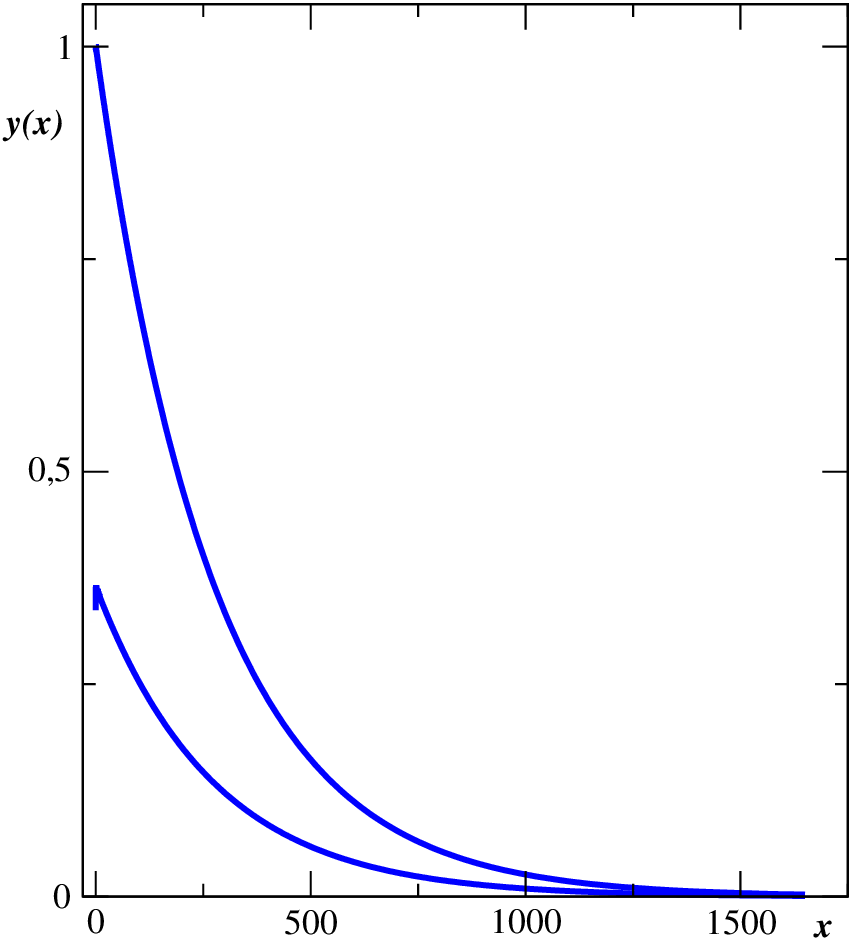}
$~~~~~~~~~~$\includegraphics[height=6.cm,width=6cm]{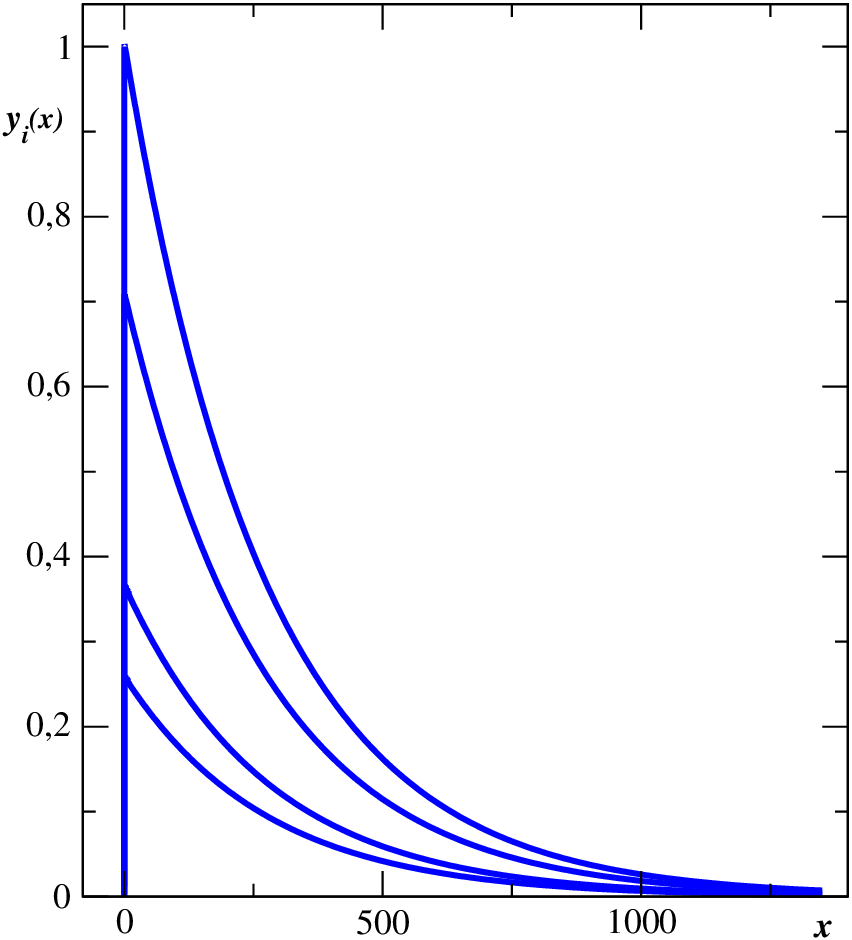}$~~~$
\end{center}

\medskip

\begin{footnotesize}
\noindent
FIG. 4 Left. The components of the ground state of Parapositronium:
$y_1(x)$ (upper) and $- 200\,y_3(x)$ (lower). 

\noindent \phantom{FIG. 4} Right.
The four components of the ground state wave function of
Orthopositronium: from top to bottom 

\noindent \phantom{FIG. 4} $-\,y_4(x)$, $-\,y_3(x)$, 
$100\,y_2(x)$ and $-100\,y_1(x)$. 

\end{footnotesize}

\medskip\medskip

\subsection{The non relativistic limit}
\label{nrl}

In order to have a better insight into the solutions of
the Para- and Orthopositronium ground states, we find it useful to look
at the Schr\"odinger limit of the relativistic system and to show how
the usual equations for the Coulomb problem are recovered.

In addition to the obvious substitutions $\mu =0$ and $M=2\,m_e$, 
the non relativistic limit on the system 
(\ref{sistema_mat} -\ref{sysmat})  needs the use of the 
``atomic parameters". This means that we have to 
adopt the following rescaling for the independent variable and for the
eigenvalue:
\begin{eqnarray}
r={\displaystyle 
\frac {2\,z}{m_e\,\alpha}}\, ,
\qquad\lambda =2\,m_e\,(1 + {\displaystyle \frac {\alpha ^{2}\,w}{4}} ) \,. 
\label{atomic_variables}
\end{eqnarray}

As already noted in (\ref{evenPsi}), when dealing with the even 
problem and assuming the coefficient functions (\ref{equaUVW_C}),
the system (\ref{sistema_mat} -\ref{sysmat})
decouples in two separate subsystems, the first one
for $y_1(z)$ and $y_3(z)$, the second for $y_2(z)$ and $y_4(z)$.
Thus, isolating $\,y_3(z)$, 
\begin{eqnarray}
{y_3}(z)={\displaystyle \frac {2\,\alpha \,z}{4\,z + z\,\alpha ^{2}\,w + \alpha ^{
2}}}\,\,{\displaystyle\frac {d}{dz}}\,{y_1}(z)\,,
\end{eqnarray}
substituting it into the equation for $\,y_1(z)$ and taking the limit 
$\alpha\rightarrow 0$, we find 
\begin{eqnarray}
{\displaystyle\frac {d^{2}}{dz^{2}}}\,{y_1}(z) + {\displaystyle 
\frac {2}{z}} \,{\displaystyle \frac {d}{dz}}\,{y_1}(z) + 
 \,\Bigl(  2\,w + {\displaystyle 
\frac {2}{z}} - 
{\displaystyle \frac {j\,(j + 1)}{z^{2}}}  \Bigr)\,{y_1}(z)=0\,,
\end{eqnarray}
in which we recognize the usual $\ell=j$ radial Coulomb equation,
that gives the well known exponential solution 
${y_1}(z)=\exp(-z)$, when  $j=0$ and $w=-1/2$. 
The analogous
treatment for the second subsystem leads exactly to the same result.

For the odd case, assuming the coefficients (\ref{equaUVW_odd}), the
system resulting from (\ref{sistema_mat} -\ref{sysmat})
does not decouple: as a consequence the treatment is just a
little bit technically more complicated. 
In order to determine the non relativistic limit we have to look for the
equations satisfied by the ``large" components of the wave function.
As shown in Fig. 4, they are $y_3(z)$ and $y_4(z)$.
Therefore, after changing to the atomic
variables, we isolate $y_1(z)$ and $y_2(z)$ from the last two equations
of the system and we substitute the results in the first two ones. 
Taking the limit for $\alpha\rightarrow 0$, we get the 
following system of two coupled second order differential equations:
\begin{eqnarray}
{}& {\displaystyle\frac {d^{2}}{dz^{2}}}\,{y_3}(z) + {\displaystyle 
\frac {2}{z}}\,{\displaystyle\frac {d}{dz}}\,{y_3}(z)  + 
\Bigl(\,2\,w + {\displaystyle \frac 2{z^{2}}}-{\displaystyle \frac 
{j(j+1)+2}{z^{2}}}\,\Bigr)\,
\,{y_3}(z)  + {\displaystyle \frac {2}{z^{2}}}\,
\sqrt{j(j +1)}\,
\,{y_4}(z)=0\spazio{1.6}\cr
{}& {\displaystyle\frac {d^{2}}{dz^{2}}}\,{y_4}(z) + {\displaystyle 
\frac {2}{z}}\,{\displaystyle\frac {d}{dz}}\,{y_4}(z)  + 
\Bigl(\,2\,w + {\displaystyle \frac 2{z^{2}}}-{\displaystyle \frac 
{j(j+1)}{z^{2}}}\,\Bigr)\,
\,{y_4}(z)  + {\displaystyle \frac {2}{z^{2}}}\,
\sqrt{j(j +1)}\,
\,{y_3}(z)=0
\label{SchrPosiOdd}
\end{eqnarray}
The differential operators of both equations of (\ref{SchrPosiOdd})
coincide. The matrix of the non-differentiated terms
\begin{eqnarray}
\left[ 
{\begin{array}{cc}
 2\,w + {\displaystyle \frac 2z} - {\displaystyle \frac{j(j+1) +2}{z^2}}
   & 2\,\sqrt{j(j + 1)}
 \spazio{1.4}\cr
 2\,\sqrt{j(j + 1)} &
2\,w + {\displaystyle \frac 2z} - {\displaystyle \frac{j(j+1)}{z^2}}
\end{array}}
 \right] 
\end{eqnarray}
is diagonalized by the constant matrix  
\begin{eqnarray}
T\,=\,\left[ 
{\begin{array}{cc}
 \sqrt{j+1} &  \sqrt{j}
 \spazio{1.}\cr
 - \sqrt{j} & \sqrt{j+1}
\end{array}}
 \right]\,. 
\end{eqnarray}
Therefore, letting $~{}^t\!(\varphi_1(z),\varphi_2(z))=
(T^{-1})\,\,{}^t\!(y_1(z),y_2(z))~$ we obtain the two decoupled equations
\begin{eqnarray}
{\displaystyle\frac {d^{2}}{dz^{2}}}\,{\varphi_1}(z) + {\displaystyle 
\frac {2}{z}} \,{\displaystyle \frac {d}{dz}}\,{\varphi_1}(z) + 
 \,\Bigl(  2\,w + {\displaystyle 
\frac {2}{z}} - 
{\displaystyle \frac {(j + 1)\,(j + 2)}{z^{2}}}  
\Bigr)\,{\varphi_1}(z)=0\,,\spazio{1.6}\cr
{\displaystyle\frac {d^{2}}{dz^{2}}}\,{\varphi_2}(z) + {\displaystyle 
\frac {2}{z}} \,{\displaystyle \frac {d}{dz}}\,{\varphi_2}(z) + 
 \,\Bigl(  2\,w + {\displaystyle 
\frac {2}{z}} - 
{\displaystyle \frac {j\,(j - 1)}{z^{2}}}  \Bigr)\,{\varphi_2}(z)=0\,,
\end{eqnarray}
corresponding to the radial Coulomb equations with $\ell=j+1$ and 
$\ell=j-1$.

\subsection{Some considerations on the hyperfine shifts}

Comparing the limiting classical wave functions with the results
reported in Fig. 4, it appears that the relativistic corrections 
become weaker and weaker for large $z$, as expected by obvious
physical reasons. Near zero the bare Coulomb potential is no more
sufficient for an accurate description of the system and, as observed
also in case of the Lamb shift, the radiative corrections
due to the self energy and to the vacuum polarization
should be accounted for. The length that, on a physical basis, can be
assumed as a scale for the relevance of the mentioned relativistic 
effects is the  Compton wavelength of the electron 
$\lambda_{\mathrm{e}}$: it is interesting to remark
that the sharp maximum of all the
components of the Orthopositronium occurs approximately at
$(\sqrt{2}/2)\,\lambda_{\mathrm{e}}$.

These considerations can be brought to bear to a possible evaluation of
the hyperfine shift of the ground states of the Para- and
Orthopositronium. Indeed the usual and very successful
derivation of the shifts of the two ground levels is done by a
perturbative method in a non relativistic approximation.
Of course, in view of the discussion of the previous subsection,
we are able to obtain exactly the
same results if we adopt the non relativistic approximation and the
same expressions for the spin-spin interaction and for the 
annihilation energy term. The latter are derived as effective
potentials from a low momentum expansion of the 
one-photon exchange Feynman diagram, \cite{LL}, 
and the value of the hyperfine shifts of the ground levels 
are produced by the sum of two different contributions. 
The first one is the average of 
$\,4\pi\mu_0^2\,(7{\bf S}^2/3-2)\,\delta(\mathbf{r})\,$ and therefore  
implies the evaluation of the classical
Coulomb wave function at zero, as in the
original Fermi calculation \cite{EF}. This approach cannot be directly
used with the wave function we have produced, since the relativistic
effects naturally
lead to a drastic modification of their behavior just in the origin. 
The averaging over the angles of the other term,
$\,6\mu_0^2\,(({\bf S}{\bf r})({\bf S}{\bf r})/r^5-{\bf S}^2/3r^3)\,,$ 
giving a rigorously vanishing value in the non relativistic limit, 
is now negligible only for for large values of $r$ as expected, but
it is divergent in the origin, even after the radial measure
$r^2$ has been accounted for. One could try to determine a
phenomenological value of the radial coordinate where to put a kind of
cut-off both for the evaluation of the wave functions and for the
integrations as well. However, in order to use the relativistic quantum
mechanics for calculating the hyperfine splitting, we believe it would
be much more appealing to deduce an
effective interaction potential without assuming a low momentum
approximation and including radiative corrections. These ideas will be
developed elsewhere.

\bigskip\bigskip

%
%

\vfill\break


\begin{thebibliography}{999}

\bigskip

\bibitem{Kr} H. Kragh, {Am. J. Phys}
{\bf 52}, 1024, (1984).

\bibitem{Jun} K.P. Jungmann, preprint physics/9809020, (1998).

\bibitem{HHP} R. H\"ackl, V. Hund and H. Pilkuhn, 
{Phys. Rev} {\bf A 57}, 3268, (1998).

\bibitem{MCK} M. Mangin-Brinet, J. Carbonell and V.A. Karmanov, 
{Phys. Rev} {\bf C 68}, 055203, (2003).

\bibitem{CL} H. Crater and B. Liu, {Phys. Rev.} {\bf C 67}, 024001, 
(2003).

\bibitem{CVA} H. Crater and P. Van Alstine, {Phys. Rev.} {\bf D 70}, 
034026, (2004).

\bibitem{SZ} E.V. Shuryak, I. Zahed,  preprint hep-ph/0403127, (2004).

\bibitem{Chi} R.W. Childers, {Phys. Rev.} {\bf D 26}, 2902, (1982).

\bibitem{SSM} T.C. Scott, J. Shertzer and R.A. Moore, 
{Phys. Rev} {\bf A 45}, 4393, (1992).

\bibitem{Dare} J.W. Darewych and L. Di Leo, {J. Phys. A}, {\bf 29},
6817, (1996); J.W. Darewych and A. Duvirak, preprint nucl-th 0204006,
(2002).

\bibitem{IG} S. Goffrey and N. Isgur, {Phys. Rev.} {\bf D 32}, 189, 
(1985).

\bibitem{RKM} R. Ricken, M. Koll, B.Ch. Metsch and H.R. Petry, 
{Eur. Phys. J.} {\bf A 9}, 221,  (2000).

\bibitem{AAVV} W.N. Polyzou, {J. Math. Phys} {\bf 43}, 6024,
(2002).

\bibitem{GS} R. Giachetti and E. Sorace, {Nuovo Cimento}
{\bf 63B}, 666, (1981). 

\bibitem{ST} E. Sorace and M. Tarlini, {Nuovo Cimento}
{\bf 71B}, 98, (1982) and {Lett. Nuovo Cimento} {\bf 35}, 1, (1982),
E. Sorace, {Ann. Henri Poincar\'e}, {\bf 3}, 659 (2002) .

\bibitem{GRR} R. Giachetti, R.Ragionieri and R. Ricci, {J. Diff.Geom.}
{\bf 16}, 297, (1981); R. Giachetti and R. Ricci, {Adv.
Math.} {\bf 62}, 84, (1986). For quantization of anticommuting variables
see references in A. Barducci and R. Giachetti, {J. Phys. A: Math. Gen}
{\bf 36}, 8129 (2003). 

\bibitem{RG} R. Giachetti, {\it Hamiltonian systems with symmetry: 
an introduction}, {Riv. Nuovo Cimento} {\bf 4}, 1,
(1981).

\bibitem{BT} B. Bakamjian and L. H. Thomas, {Phys. Rev.} {\bf 92}, 1300,
(1953).

\bibitem{Landau} L. Landau and L. Lifschitz,  {\it Th\'eorie des Champs},
Edition MIR, (Moscou,1970).

\bibitem{LoLu} G. Longhi and L. Lusanna,  {Phys. Rev.} {\bf D 34}, 3707, 
(1986).

\bibitem{VW} S. Vugalter and T. Wiedl, {Ann. Henri Poincar\'e}  
{\bf 4}, 301 (2003).

\bibitem{Ince} E.L. Ince,  {\it Ordinary differential equations},
Dover Publications Inc., (New York,1956). 

\bibitem{LL} L. Landau and L. Lifschitz,  {\it Th\'eorie quantique
relativiste},
Edition MIR, (Moscou,1972); B. Thaller {\it The Dirac Equation},
(Springer Verlag, Berlin, 1992); F.J. Yndurain 
{\it Relativistic Quantum Mechanics and Introduction to Field Theory}, 
(Springer Verlag, Berlin, 1996).

\bibitem{DS} N. Dunford and J. Schwartz,  {\it Linear Operators, 
Part II}, Wiley Interscience Publishers, (New York,1963).

\bibitem{Breit} G. Breit, {Phys. Rev.} {\bf 36}, 383 (1930).

\bibitem{FFT} G. Feldman, T. Fulton and J. Townsend, {Phys. Rev.} 
{\bf A8}, 1149 (1973)

\bibitem{Cope} F.T. Cope, {Amer. J. Math.} {\bf 56}, 411 (1934) and
{\bf 58}, 130 (1936).

\bibitem{EF} E. Fermi, {Z. Phys.} {\bf 60}, 320 (1930).


\end{thebibliography}
\end{document}